\documentclass[twocolumn,txfonts,showpacs,nofootinbib,superscriptaddress,preprintnumbers,amsmath,amssymb,epsfig,color]{aa}
\usepackage[colorlinks=true,pdfstartview=FitV,linkcolor=blue,citecolor=blue,urlcolor=blue]{hyperref}  

\usepackage[usenames]{color}
\usepackage{graphicx}%
\usepackage{units, subfigure,amssymb,amsmath}
\usepackage[latin1]{inputenc}
\usepackage{txfonts} 
\newcommand{\captionsize}{\small}

\newcommand{\AMS}{\textsf{AMS}}
\newcommand{\Hyd}{\textsf{H}}
\newcommand{\He}{\textsf{He}}
\newcommand{\Li}{\textsf{Li}}
\newcommand{\Be}{\textsf{Be}}
\newcommand{\B}{\textsf{B}}
\newcommand{\C}{\textsf{C}}
\newcommand{\N}{\textsf{N}}
\newcommand{\Oxy}{\textsf{O}}

\newcommand{\F}{\textsf{F}}
\newcommand{\Ne}{\textsf{Ne}}
\newcommand{\Al}{\textsf{Al}}
\newcommand{\Mg}{\textsf{Mg}}
\newcommand{\Si}{\textsf{Si}}

\newcommand{\Ti}{\textsf{Ti}}
\newcommand{\Fe}{\textsf{Fe}}

\newcommand{\BC}{\textsf{B/C}}
\newcommand{\pbarp}{\ensuremath{\overline{\textsf{p}}/\textsf{p}}}

\usepackage{natbib}
\def\Journal#1#2#3#4{{#4}, {#1}, {#2}, #3}
\newcommand{\etal}{et al.}

\def\AA{A$\&$A}
\def\ApJ{ApJ}
\def\MNRAS{MNRAS}
\def\PRL{PRL}
\def\PRD{PRD}
\def\PRC{PRL}
\def\APP{A.Ph} 
\def\AdSpRes{Adv. Sp. Res.}

\begin{document}

\title{Secondary cosmic-ray nuclei from supernova remnants   
  \\and constraints on the propagation parameters}           

\author{Nicola Tomassetti\inst{1} 
  \and   Fiorenza Donato\inst{2}
}

\offprints{{\\\tt nicola.tomassetti@pg.infn.it; fiorenza.donato@unito.it}}

\institute{
  INFN -- Sezione di Perugia, 06122 Perugia, Italy
  \and Physics Department, Torino University and INFN, 10125 Torino, Italy 
} 
\date{Received 6 February 2012 / Accepted 23 March 2012}

\abstract       
{
The secondary-to-primary boron-to-carbon (\BC) ratio is widely used to study the cosmic-ray (CR) propagation processes 
in the Galaxy. It is usually assumed that secondary nuclei such as
\textsf{Li-Be-B} are generated entirely by collisions of heavier CR nuclei with the interstellar medium (ISM). 
}
{
We study the CR propagation under a scenario where secondary nuclei can also be produced 
or accelerated by Galactic sources. We consider the processes of hadronic interactions inside 
supernova remnants (SNRs) and the re-acceleration of background CRs in strong shocks. We investigate 
their impact in the propagation parameter determination within present and future data.  
}
{
Analytical calculations are performed in the frameworks of the diffusive shock 
acceleration theory and the diffusive halo model of CR transport. 
Statistical analyses are performed to determine the  propagation parameters 
and their uncertainty bounds using existing data on the \BC{} ratio, as well as the 
simulated data expected from the \AMS-02{} experiment.
}
{
The spectra of \Li-\Be-\B{} nuclei emitted from SNRs are harder than those due to CR collisions with the ISM. 
The secondary-to-primary ratios flatten significantly at $\sim$\,TeV/n energies, both from spallation and 
re-acceleration in the sources. 
The two mechanisms are complementary to each other and depend on the properties of the local ISM around 
the expanding remnants. The secondary production in SNRs is significant for dense background 
media, $n_{1}\gtrsim 1$\,cm$^{-3}$, while the amount of re-accelerated
CRs is relevant to SNRs 
expanding into rarefied media, $n_{1}\lesssim 0.1$\,cm$^{-3}$. Owing these effects, the diffusion parameter $\delta$ 
may be underestimated  by a factor of $\sim$\,5--15\%. Our estimations indicate that an experiment of the \AMS-02{} caliber
can constrain the key propagation parameters, while breaking the source-transport degeneracy for a wide class of \BC-consistent models. 
}
{
Given the precision of the data expected from ongoing experiments, the SNR 
production/acceleration of secondary nuclei should be considered, if any, to prevent a 
possible mis-determination of the CR transport parameters. 
}

\authorrunning{ Tomassetti \& Donato}
\titlerunning{Secondary CRs from SNRs and propagation parameters} 
\maketitle

\section{Introduction}    
\label{Sec::Introduction} 

The problems of the origin and propagation of the charged cosmic rays (CRs) in the Galaxy are among 
the major topics of resarch in
modern astrophysics. It is generally accepted that \textit{primary} CR 
nuclei such as \Hyd, \He, \C, \N, and \Oxy, are accelerated in supernova remnants (SNRs) via diffusive shock 
acceleration (DSA) mechanisms, that produce power-law momentum
spectra \citep{Drury1983}. At relativistic energies, $S \propto p^{-\nu} \sim E^{-\nu}$.
After being accelerated, CRs are released in the circumstellar environment, where they diffuse through 
the turbulent magnetic fields and interact with interstellar matter (ISM) \citep{Strong2007}. 
Owing to diffusion, CRs stream out from the Galaxy on a
characteristic timescale $\tau_{\rm esc}\propto E^{-\delta}$. 
The spectrum of primary CR nuclei predicted at Earth is therefore $N_{p} \sim S \tau_{\rm esc}$.
The collisions of these nuclei with interstellar gas are believed to be the mechanism producing the
\textit{secondary} CR nuclei, such as \Li, \Be, and \B, which are under-abundant in the thermal ISM.
Thus, the equilibrium spectra of secondary CRs are $E^{-\delta}$ times softer than those of their progenitors.
At energies above some tenths of GeV per nucleon, where CR nuclei reach the pure diffusive regime, 
this picture predicts power-law distributions such as $N_{p}\sim E^{-\nu - \delta}$ for primary nuclei
and $N_{s}/N_{p} \sim E^{-\delta}$ for secondary-to-primary ratios at Earth. 
These trends may be straightforwardly derived from the analytical
solutions of \citet{Maurin2001}.
Present observations indicate that $\delta\sim$\,0.3--0.7 and $\nu\sim$\,2.0--2.4. 
The bulk of the data is collected at $E \lesssim$\,10\,GeV\,nucleon$^{-1}$,
where the CR spectra are shaped by additional effects such as diffusive reacceleration,
galactic wind convection, energy losses, and solar modulation. Since there is no firm 
theoretical prediction of the key parameters associated with these effects, it is 
very difficult to distinguish each physical component using
the experimental data. 
The boron-to-carbon (\BC) ratio is the most robustly measured secondary-to-primary ratio and is
used to constrain several model parameters. 
Throughout this paper, we call \textit{secondaries} all CR nuclei
produced by hadronic interactions, independently on the place of origin.
The standard approach, hereafter \textit{reference model}, assumes that the secondary 
nuclei are absent from the CR sources. 

In this paper, we examine two mechanisms producing a source component of secondary CRs: 
(i) the fragmentation of CR nuclei inside SNRs and 
(ii) the re-acceleration by SNRs of pre-existing CR particles.
The secondary CR production inside SNRs was studied in \citet{Berezhko2003}
and subsequently reconsidered to describe the positron fraction \citep{Blasi2009, Ahlers2009}. 
Predictions of the \pbarp{} ratio \citep{BlasiSerpico2009,Fujita2009} and \BC{} ratio 
\citep{Mertsch2009,Thoudam2011} have also been investigated. 
An interesting aspect of this mechanism is that, if the secondary fragments start the DSA, 
the secondary-to-primary ratios must eventually increase. 
Similarly, the re-acceleration of background CRs interacting with the expanding SNR shells 
may induce a significant transformation of their spectra 
at high energies \citep{Berezhko2003,Wandel1987}.  
In particular, the re-acceleration redistributes the spectrum of secondary nuclei to a spectrum $S\sim~E^{-\nu}$.
The main feature of both mechanisms is that they produce harder spectra of secondary nuclei 
than in the case of their standard production from primary CR
collisions in the ISM. 
These source components of secondary CRs may become relevant at $\sim$\,TeV energies.
Thus, disregarding these effects may lead to a mis-determination of the CR transport parameters.
The aim of this paper is to examine their impact on the CR propagation physics. 
This task requires a description of the CR acceleration processes in SNRs and 
their interstellar propagation. In this work, we use fully analytical calculations in 
the frameworks of the linear DSA theory and the diffusion halo model (DHM) 
of CR transport. 
In Sect.\,\S\ref{Sec::SNRAcceleration}, we present the DSA calculations for CR nuclei, 
including standard injection from the thermal ISM, hadronic interactions, and re-acceleration. 
Sect.\,\S\ref{Sec::GalacticPropagation} outlines the basic elements of the DHM galactic propagation.
In Sect.\,\S\ref{Sec::Results}, we show our model predictions for the 
CR spectra and ratios at Earth, and study the impact of the secondary source components 
on the determination of the CR transport parameters. 
In Sect.\,\S\ref{Sec::AMSPhysics}, we present our estimates for the Alpha Magnetic 
Spectrometer (\AMS). We present our conclusions in Sect.\,\S\ref{Sec::ConclusionAndDiscussion}.

\section{Acceleration in SNR shock waves }  
\label{Sec::SNRAcceleration}                

We compute the spectrum of CR ions accelerated in SNRs using the 
DSA theory \citep{Drury1983}, including the loss and source terms
and both the production and acceleration of secondary fragments. 
Our derivation is formally similar to that in \citet{Morlino2011}, but the
physical problem is the same to that treated in \citet{Mertsch2009}. 
Within this formalism, we also compute the re-acceleration of
pre-existing CR particles. 
%

\subsection{DSA Calculations}  
\label{Sec::AccelerationDSA}   

We consider the case of plane shock geometry and a test-particle approximation, \textit{i.e.}, 
we ignore the feedback of the CR pressure on the shock
dynamics. The shock front is in 
its rest-frame at $x = 0$. The un-shocked upstream plasma flows in from $x < 0$ at speed 
$u_{1}$ (density $n_{1}$) and the shocked downstream plasma flows out to $x > 0$ at speed 
$u_{2}$ (density $n_{2}$). These quantities are related by the compression ratio 
$r=u_{1}/u_{2} = n_{2}/n_{1}$. 
The particle spectra are described by the phase space density $f(p,x)$. 
The equation that describes the diffusive transport and convection at the shock for 
a $j$--type nucleus (charge $Z_{j}$ and mass number $A_{j}$) is given by:
\begin{equation} \label{Eq::DiffusionDSA}
 u \frac{\partial f_j}{\partial x} = D_{j} \frac{\partial^{2}
 f_j}{\partial x^{2}} + 
 \frac{1}{3}\frac{du}{dx}p\frac{\partial f_j}{\partial p} 
 -\Gamma^{\rm inel}_j{f_j}  + Q_j \,,
\end{equation}
where $D_{j}(p)$ is the diffusion coefficient near the SNR shock, $u$ is the fluid velocity, 
$\Gamma^{\rm inel}_{j} = \beta_{j} c n \sigma^{\rm inel}_{j}$ is the total destruction rate for fragmentation (see Sect.\,\S\ref{Sec::SecondaryNuclei}),
$\sigma^{\rm inel}_{j}$ is the cross-section for the process, and $Q_j(x,p)$ represents the source term.
Solutions of Eq.\,\ref{Eq::DiffusionDSA} can be found, separately, in the regions upstream 
($x<0$) and downstream ($x>0$) of the shock front, by requiring that $\partial f/\partial x=0$ for 
$x \rightarrow \mp \infty$. 
We drop the label $j$ characterizing the nuclear species, and make use of the subscript $i=1$ ($i=2$) 
to indicate the quantities in the upstream (downstream) region. We define the quantities
\begin{equation}\label{Eq::DefLambdaKappa}
  \lambda_{i} = \frac{u_{i}}{D_{i}}( \Lambda_{i} -1  ) \qquad\qquad
  \kappa_{i} = \frac{u_{i}}{D_{i}}( \Lambda_{i} +1  ) \,,
\end{equation}
where $\Lambda_{i} = \sqrt{1 + 4D_{i}\Gamma^{\rm inel}_{i}/u_{i}^{2}}$. The solution can 
be expressed in the form
\begin{equation}\label{Eq::FullSolutionUSDS}
f(x,p) =
  \begin{cases}
    f_{0}(p) e^{-\frac{1}{2}\kappa_{1}x}  - \frac{U_{1} + V_{1} + W_{1}}{u_{1} \Lambda_{1}} & (x<0) \,,\\
    \\
    f_{0}(p) e^{+\frac{1}{2}\lambda_{2}x} + \frac{U_{2} + V_{2} + W_{2}}{u_{2} \Lambda_{2}} & (x>0)  \,,
  \end{cases}
\end{equation}
where the downstream integral terms $U_{2}$, $V_{2}$, and $W_{2}$ are given by:
\begin{equation}\label{Eq::Integrals}
  \begin{aligned}
  U_{2}(x,p) &= +\int_{x}^{+\infty} Q_{2}(x',p) e^{\frac{1}{2}\kappa_{2}(x-x')} dx' \,,\\
  V_{2}(x,p) &= +\int_{0}^{x} Q_{2}(x',p) e^{-\frac{1}{2}\lambda_{2}(x-x')} dx' \,,\\
  W_{2}(x,p) &= -\int_{0}^{+\infty} Q_{2}(x',p) e^{-\frac{1}{2}(\lambda_{2} x + \kappa_{2}x')} dx' \,.
  \end{aligned}
\end{equation}
In the upstream region, $U_{1}$, $V_{1}$, and $W_{1}$ are still given by Eq.\,\ref{Eq::Integrals} 
after performing the substitutions $1\rightarrow 2$, $\kappa_{2} \rightarrow -\lambda_{1}$,  
$\lambda_{2} \rightarrow -\kappa_{1}$, and $\infty \rightarrow -\infty$. 
The distribution function at the shock position, $f_{0}$, is determined by the matching conditions at $x=0$. 
We integrate Eq.\,\ref{Eq::DiffusionDSA} in a thin region across the shock front. 
Assuming that $D\equiv D_{1} = D_{2}$, we find the equation for $f_{0}$ 
\begin{equation}\label{Eq::DiffEquationAtShockFront}
p \frac{\partial f_{0}}{\partial p} = -\alpha f_{0}(p) - \alpha j(p) + \frac{\alpha}{u_{1}}G(p) \,,
\end{equation}
where $\alpha = 3u_{1}/(u_{1}-u_{2})$ is the known DSA spectral index. 
The term $G$ denotes the sum of the upstream and downstream source integrals 
\begin{equation}
  G(p) = 
  \int_{-\infty}^{0}Q_{1} e^{\frac{1}{2}\lambda_{1}x'} dx' + 
  \int_{0}^{\infty}Q_{2} e^{-\frac{1}{2}\kappa_{2}x'} dx' \,.
\end{equation}
The function $j(p)$ is linked to the destruction term $\Gamma^{\rm inel}$. It is defined as 
\begin{equation}\label{Eq::JFunction}
  j(p) =  \frac{1}{2} ( \Lambda_{1} - 1 ) + \frac{1}{2r} ( \Lambda_{2} - 1 ) \,.
\end{equation}
After defining the function
\begin{equation}\label{Eq::Chi}
\chi(p,p^{\prime}) = \alpha \int_{p^{\prime}}^{p} \frac{j(p^{\prime\prime})}{p^{\prime\prime}} dp^{\prime\prime} \,, 
\end{equation}
the solution of Eq.\,\ref{Eq::DiffEquationAtShockFront} can be expressed in the simple form 
\begin{equation}\label{Eq::FullSolutionAtShockFront}
f_{0}(p) = \alpha \int_{0}^{p} \left( \frac{p^{\prime}}{p}
\right)^{\alpha} \frac{G(p^{\prime})}{u_{1}} e^{-\chi(p,p^{\prime})}
\frac{dp^{\prime}}{p^{\prime}} \,.
\end{equation}
From Eq.\,\ref{Eq::FullSolutionUSDS}, one recovers the standard DSA
solution by setting 
$\Gamma^{\rm inel}=0$ (no interactions) and assuming that the injection occurs only at the shock 
front ($U_{i}=V_{i}=W_{i}=0$): 
one finds that $f_{2} = f_{0}$ and $f_{1} = f_{0} e^{u_{1}x/D_{1}}$, while 
Eq.\,\ref{Eq::FullSolutionAtShockFront} gives a spectrum $p^{-\alpha}$, 
provided that the source term $G(p)$ is softer than $p^{-\alpha}$ (see Sect.\,\S\ref{Sec::ReAcceleratedNuclei}).
Some simplifications can be made by analyzing the timescales of the problem. 
The DSA acceleration rate for particles of momentum $p$ 
in a stationary shock is $\Gamma^{\rm acc} \sim \frac{u_{1}^{2}}{20\,D}$,
where we assumed Bohm diffusion ($D\propto p$) and strong shocks ($r\approx 4$). 
For a SNR of age $\tau_{\rm snr}$, the condition $\Gamma^{\rm acc}\equiv\tau^{-1}_{\rm snr}$ defines 
the maximum momentum $p^{\rm max}$ attainable by DSA. 
In the presence of hadronic interactions, the requirement 
$\Gamma^{\rm inel} \ll \Gamma^{\rm acc}$ must be fulfilled. 
These relations imply that 
$20\,\Gamma^{\rm inel} D /u^{2} \ll 1$ and $x\Gamma^{\rm inel}/u \ll 1$ at all the energies considered. 
Under these conditions, we can linearly expand $\Lambda \approx 1 + 2\Gamma^{\rm inel} D/u^{2}$, 
so that $\lambda \approx 2\Gamma^{\rm inel}/u$ and $\kappa \approx 2u/D + 2\Gamma^{\rm inel}/u$. 
The exponential terms of Eq.\,\ref{Eq::FullSolutionUSDS} and Eq.\,\ref{Eq::Integrals} 
can also be expanded as $e^{\frac{1}{2} \lambda x} \approx 1 - \frac{\Gamma^{\rm inel}}{u}x$ and  
$e^{\frac{1}{2} \kappa x} \approx e^{u/D} \left( 1 + \frac{\Gamma^{\rm inel}}{u}x\right)$. 
Thus, the function $j(p)$ of Eq.\,\ref{Eq::JFunction} is given by
\begin{equation}\label{Eq::JFunctionApprox}
  j(p) \approx \alpha (1+r^{2}) \frac{\Gamma^{\rm inel}_{1} D(p)}{u_{1}^{2}} \,,
\end{equation}
and the integral of Eq.\,\ref{Eq::Chi} by
\begin{equation}\label{Eq::ChiApprox}
  \chi(p,p^{\prime}) \approx  \alpha (1+r^{2}) \frac{\Gamma^{\rm inel}_{1}}{u_{1}^{2}} [ D(p) - D(p^{\prime}) ] \,,
\end{equation}
which recovers the expression of \citet{Mertsch2009}. 
Below we present the DSA solutions for primary nuclei (injected at the shock), 
their secondary fragments (generated in the SNR environment), 
and pre-existing CR particles that undergo re-acceleration.

\begin{figure*}[!ht]
\begin{center}
\includegraphics[width=1.60\columnwidth]{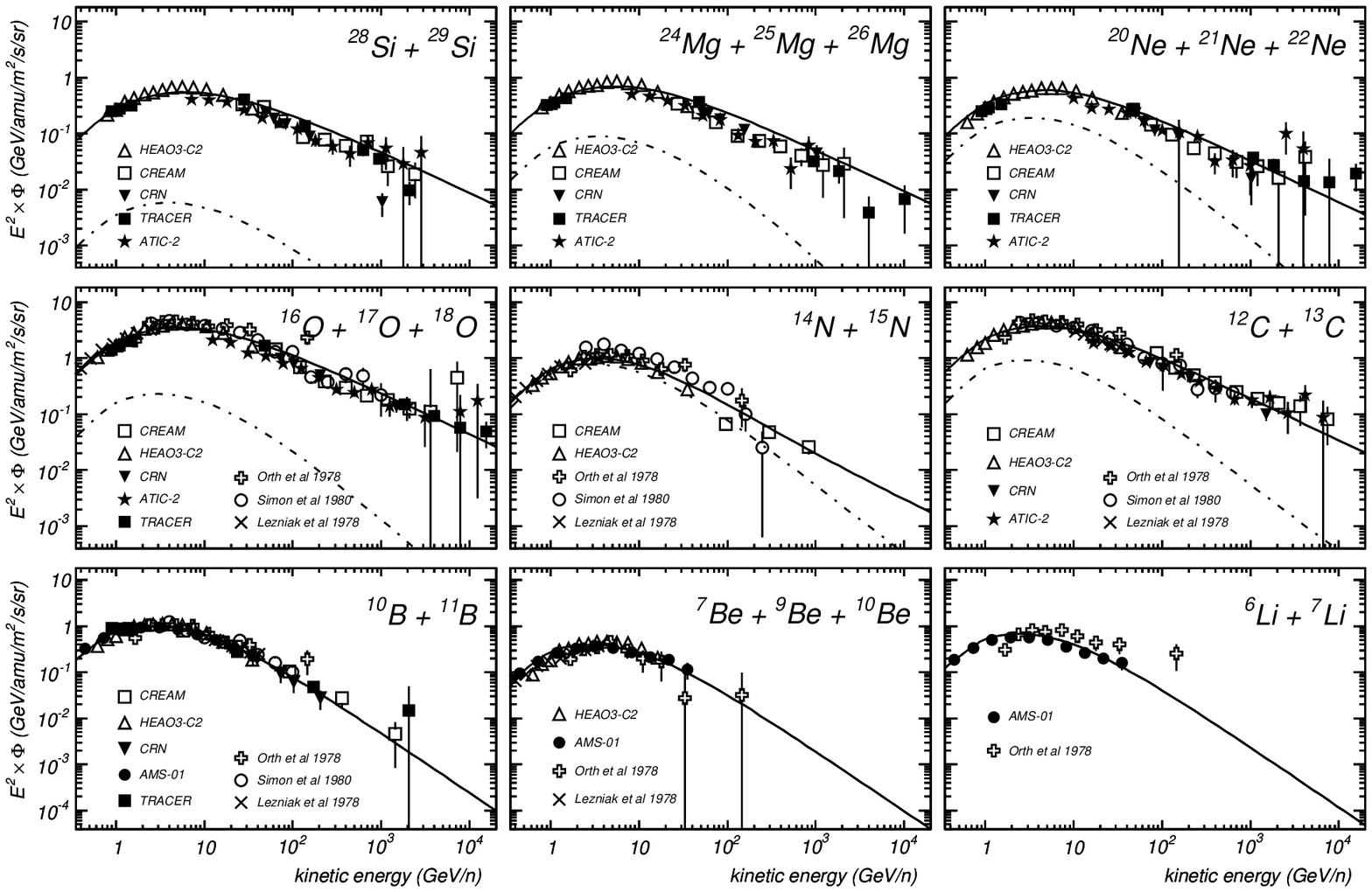}
\caption{\captionsize 
  Energy spectra of the CR elements \Li, \Be, \B, \C, \N, \Oxy, \Ne, \Mg{} and \Si.
  The solid lines represent the \textit{reference model} prediction.
  The dashed lines indicate the secondary CR component arising by
  collisions of heavier nuclei in the ISM. 
  The model parameters are listed in Table\,\ref{Tab::Parameters}.
  Data are from HEAO3-C2 \citep{Engelmann1990}, CREAM \citep{Ahn2009}, \AMS-01 \citep{Aguilar2010},
  TRACER \citep{Ave2008,Obermeier2011}, ATIC-2 \citep{Panov2009}, CRN~\citep{Muller1991}, \citet{Simon1980}, 
  \citet{Lezniak1978} and \citet{Orth1978}. The \Li-\Be-\B{} data from CREAM and \AMS-01 are combined with our 
  model to obtain the spectra from their secondary-to-primary ratios. 
\label{Fig::ccAllSpectra} 
}
\end{center}
\end{figure*}

\subsection{Acceleration of primary nuclei} 
\label{Sec::PrimaryNuclei}                  

The injection of ambient particles is assumed to occur immediately
upstream from 
the shock at momentum $p^{\rm inj}$. The source term for primary nuclei is
\begin{equation} \label{Eq::Injection}
 Q^{\rm pri}(x,p) =  Y \, \delta(x) \, \delta(p-p^{\rm inj}) \,.
\end{equation}
Particles can be injected only when their \textit{Larmor radius} is large enough to
cross the shock thickness. Thus, we assume a reference injection rigidity for 
all nuclei, $R^{\rm inj}$, so that $p^{\rm inj}= Z R^{\rm inj}$.
The constant $Y$ reflects the particle abundances in the ISM and their injection efficiencies. 
In this work, they are determined by the data. The phase space density profile is given by
\begin{equation}\label{Eq::DSAStandardPlusBreakup}
  f(x,p) \approx
  \begin{cases}
    f_{0}(p) e^{u_{1}x/D(p)}  & (x<0) \,, \\
    f_{0}(p) \left( 1 - \frac{\Gamma^{\rm inel}_{2}}{u_{2}} x \right) & (x>0) \,.
  \end{cases}
\end{equation}
The upstream profile, $\sim e^{u_{1}x/D}$, indicates that the plasma is confined near the shock within a 
typical distance $\sim D/u_{1}$. Owing to advection, the particles are accumulated in the downstream region, 
where destruction processes give rise to the term $\frac{\Gamma^{\rm inel}_{2}}{u_{2}}x$.
The momentum spectrum at the shock position is $f_{0} \propto e^{-\chi} p^{-\alpha}$, 
which is the known DSA power-law behavior times an exponential factor, $\sim e^{-\chi}$,  
given by $\chi \approx \alpha \Gamma^{\rm inel}/\Gamma^{\rm acc}$.
The condition $\Gamma^{\rm acc} \ll \tau^{-1}_{\rm snr}$ implies that $\chi \lesssim 1$.

\subsection{Production and acceleration of secondary nuclei} 
\label{Sec::SecondaryNuclei}                                 

Secondary nuclei originate in the SNR environment from the spallation of heavier nuclei on the background medium. 
The source term for a $j$-type CR species arises from the sum of all heavier $k$-type nuclei, 
$Q_{j}^{\rm sec} = \sum_{k>j}Q_{kj}^{\rm frag}$. Each partial
contribution is defined to be: 
\begin{equation} \label{Eq::SecSourceDef} 
 Q_{kj}^{\rm frag}(x,p) 
   = \frac{1}{4 \pi p^{2}} \int_{p^{\rm inj}}^{\infty} N_{k}(x,p^{\prime}) \Gamma^{\rm frag}_{kj}(p^{\prime}) \delta(p - \xi_{kj}p^{\prime}) dp' \,,
\end{equation}
where $N_{k}(x,p^{\prime})= 4 \pi {p^{\prime}}^{2} f_{k}(x,p^{\prime})$ is the progenitor number density
and $\Gamma^{\rm frag}_{kj} = \beta_{k} c n \sigma^{\rm frag}_{kj}$ is the $k\rightarrow j$ fragmentation rate,
which is implicitly summed over the circumstellar abundances (the hydrogen and helium components).   
We have assumed that the kinetic energy per nucleon is conserved in the process, \textit{i.e.}, 
the fragments are ejected with momentum $\xi_{kj} = A_{j}/A_{k}$ times smaller than that of their parents. 
The presence of short-lived isotopes (\textit{ghost nuclei}) such as
$^{9}$\Li{} or $^{11}$\C{} is reabsorbed in the definition of $\sigma_{kj}$,
while long-lived isotopes such as $^{10}$\Be{} or $^{26}$\Al{} (lifetime $\sim\,$1\,Myr) are considered as stable during the 
acceleration process (timescale $\tau_{\rm snr} \sim\,$10\,kyr).  
The source spatial profile takes the form of the progenitor nucleus of Eq.\,\ref{Eq::DSAStandardPlusBreakup}. 
In the upstream region, one has $Q^{\rm frag}_{jk} = q_{1,kj}e^{u_{1}x/D_{k}}$, where $D_{k}$ is the diffusion 
coefficient of the progenitor nucleus at the momentum $p/\xi_{kj}$.
For $D \propto p/Z$, one can write $D_{k}(p/\xi_{kj}) = \zeta_{kj}D_{j}(p)$, where $\zeta_{kj} = \frac{A_{j}Z_{k}}{A_{k}Z_{j}}$. 
In practice, $\zeta_{kj} \approx 1$ for all processes $k\rightarrow j$ with $Z_{j,k}>2$. 
The terms at the shock, $q_{1,jk}$ and $q_{2,jk}$, are given by
\begin{equation}
q_{i,kj}(p) = \xi_{kj}^{-3}f_{0,k}(p/\xi_{kj})\Gamma^{\rm frag}_{i,kj} \,,
\end{equation}
and the downstream solution reads
\begin{equation}\label{Eq::DownStreamSecondary}
f_{2,j}(x,p) = f_{0,j}(p) + \left[ \frac{q_{2,j}(p)}{u_{2}}- \Gamma^{\rm frag}_{2,j}f_{0,j}(p) \right] x \,,
\end{equation}
where $f_{0,j}$ is given from Eq.\,\ref{Eq::FullSolutionAtShockFront} using $G(p)$ by the expression
\begin{equation}\label{Eq::GTermSecondary}
  G_{kj}(p) = q_{1,kj}(p) \frac{D_{j}(p)}{u_{1}}\left( \zeta_{kj}^{-1} + r^{2} \right) \,.
\end{equation}  
We solve all equations starting from the heaviest element and proceeding downward in mass.
We are interested in the total contribution of SNRs to the Galactic CR population,
which we evaluate to be the integral of the downstream solution over the SNR volume left behind the shock
\begin{equation}\label{Eq::SNRVolumeIntegral}
S^{\rm dsa}_{j}(p) = 4\pi p^{2}\mathcal{R_{\rm snr}} \int_{0}^{x_{\rm max}} 4 \pi x^{2} f_{2,j}(x,p) dx \,,
\end{equation}
where $x_{\rm max}=u_{2}\tau_{\rm snr}$ and $\mathcal{R_{\rm snr}}$ is the supernovae explosion rate per unit volume in the Galaxy.

\subsection{Re-acceleration of background CR nuclei}  
\label{Sec::ReAcceleratedNuclei}                      

Together with the thermal ISM particles of Sect.\,\S\ref{Sec::PrimaryNuclei}, SNR shock waves may 
also accelerate the background CRs at equilibrium \citep{Berezhko2003,Wandel1987}.
We refer to this mechanism as the \textit{re-acceleration} of CRs in SNRs. 
For a prescribed distribution function of background CRs, $f_{j}^{\rm bg}(p)$, the DSA solution at 
the shock is simply
\begin{equation}\label{Eq::ReAccelerationAtShock}
  f_{0,j}^{\rm re}(p) = \alpha \int_{p_{j}^{\rm inj}}^{p} \left(
    \frac{p^{\prime}}{p} \right)^{\alpha} f_{j}^{\rm bg}(p^{\prime})
  \frac{dp^{\prime}}{p^{\prime}} \,.
\end{equation}
Assuming, for illustrative purposes, a power-law form $f_{j}^{\rm bg}(p) = Y_{j}(p/p_{j}^{\rm inj})^{-s}$, 
the resulting re-acceleration spectrum is
\begin{equation}
 f_{0,j}^{\rm re}(p) = \frac{\alpha}{\alpha-s} \left[ 1 - (p/p^{\rm
     inj}_{j})^{-\alpha + s} \right]f^{\rm bg}_{j}(p)  \,.
\end{equation}
Since the CR equilibrium spectrum, $f^{\rm bg}\propto p^{-s}$, is softer than the 
test-particle one ($s > \alpha$), for $p \gg p_{j}^{\rm inj}$ one obtains
\begin{equation}
f_{0,j}^{\rm re}(p)  \approx \frac{\alpha}{s-\alpha}Y_{j} \left( p/p^{\rm inj}_{j} \right)^{-\alpha} \,.
\end{equation}
That is, 
the effect of re-acceleration is to re-distribute the CR spectrum to $p^{-\alpha}$. Interestingly, 
in the opposite case ($s < \alpha$) the re-accelerated spectrum maintains its spectral shape 
$p^{-s}$, while its normalization is amplified by the factor $\alpha/(s-\alpha)$.
In our model, however, the background spectrum $f^{\rm bg}$ is computed 
as discussed in Sect.\,\S\ref{Sec::GalacticPropagation} and takes the SNR spectra as input.
Therefore, Eq.\,\ref{Eq::ReAccelerationAtShock} is an integro-differential equation 
where the DSA-mechanism is fed by its DHM-propagated solution and vice-versa.
On the other hand, the bulk of the re-accelerated CRs come from the low-energy part of the spectrum 
(below $\sim$\,10\,GV of rigidity), where the equilibrium CR spectra are fixed by the observations 
so they cannot vary too much. 
It can be safely assumed that re-accelerated CRs are a sub-dominant component of the total (integral) flux.
Hence, we proceed using an iterative method as outlined in Sect.\,\S\ref{Sec::ReAcceleratedNucleiInSNRs}.

\section{Interstellar propagation}  
\label{Sec::GalacticPropagation}    

We use a DHM to describe the CR transport and interactions in the ISM
in a two-dimensional geometry. We disregard the effects 
of energy losses, diffusive reacceleration, and convection. The Galaxy is 
modeled as a disk of half-thickness $h$, containing the gas and the CR sources. The disk is 
surrounded by a cylindrical diffusive halo of half-thickness $L$, radius $r_{\rm max}$, and zero 
matter density. The CRs diffuse into both the disk and the halo. The diffusion coefficient is taken 
to be rigidity dependent and position independent such that $K(R) = \beta K_{0}(R/R_{0})^{\delta}$.
The number density $N_{j}$ of the nucleus $j$ is a function of the kinetic energy per nucleon, 
$E$, and the position $(r,z)$. The steady-state transport equation can
be written as 
\begin{equation}\label{Eq::DHMEquation}
\left( \mathcal{W}_{j}^{\rm tot} - K_{j} \nabla^{2}  \right) N_{j} =  \mathcal{S}_{j}^{\rm tot} \,.
\end{equation}
The loss term, $\mathcal{W}_{j}^{\rm tot}$, describes the decay rate of unstable nuclei, 
$\tilde{\Gamma}_{j}^{\rm rad}=1/(\gamma_L\tau_{j})$ (where $\gamma_{L}$ is the usual Lorentz factor)
and the total destruction rate for collisions in the disk, 
$2h\delta(z)\tilde{\Gamma}_{j}^{\rm inel}$. 
The source term, $\mathcal{S}^{\rm tot}_{j}$, is the sum of contributions from SNRs (the DSA solution 
of Eq.\,\ref{Eq::SNRVolumeIntegral}), and the secondary production in the ISM from $k$-type progenitors
\begin{align}
  &\mathcal{S}_{j}^{\rm snr} = 2h\delta(z)s(r)S_{j}^{\rm dsa}(E) \label{Eq::DHMSourceSNR} \,, \\
  &\mathcal{S}_{j}^{\rm ism} = 2h\delta(z) \sum_{k>j} \left(\tilde\Gamma_{kj}^{\rm frag} + \tilde\Gamma_{kj}^{\rm rad} \right) N_{k} \label{Eq::DHMSourceIsm} \,.
\end{align}
The function $s(r)$ expresses the SNR radial distribution in the disk,
that we assume to be uniform. 
For primary CRs, ${S}^{\rm dsa}$ is normalized by the $Y$ constants of Eq.\,\ref{Eq::Injection}. 
In our study, secondary nuclei may have a non-zero $S^{\rm dsa}$ term. The term 
$\tilde\Gamma^{\rm rad}_{kj} = ( \gamma_L\tau_{kj} )^{-1}$ describe the contributions $k \rightarrow j$ 
from unstable progenitors of lifetime $\tau_{kj}$. 
In Eq.\,\ref{Eq::DHMSourceIsm}, $\tilde\Gamma_{j}^{\rm inel}= \beta_{j} c n_{\rm ism} \sigma^{\rm inel}_{j}$
and  $\tilde\Gamma_{kj}^{\rm frag}= \beta_{j} c n_{\rm ism} \sigma^{\rm frag}_{kj}$.
The conditions $N_{j}\equiv 0$ at the halo boundaries and the continuity condition across the disk completely 
characterize the solution of Eq.\,\ref{Eq::DHMEquation}. The full solution is reported in \citet{Maurin2001}. 
We again solve the transport equations for all the CR nuclei following their top--down fragmentation sequence,
plus a second iteration to account for the $^{10}$\Be$\rightarrow$$^{10}$\B{} decay. The differential fluxes as a function 
of kinetic energy per nucleon $E$ are obtained from
\begin{equation}\label{Eq::Flux}
\phi_{j}(E) \equiv \frac{dN_{j}}{d\Sigma d\Omega dt dE} = \frac{\beta
  c}{4 \pi}N_{j}(E, r_{\odot},0) \,,
\end{equation}
where the equilibrium solutions are computed at the Solar System position $(r,z)=(8.5{\rm\,kpc},0)$.
The basic DHM predictions can be seen, for illustrative purpose, in the one-dimensional limit 
$r_{max}\rightarrow\infty$. The solution for a pure primary CR is
given by
\begin{equation}\label{Eq::Primary1D}
  N_{p} \approx \frac{S^{\rm dsa}}{K/L + \tilde\Gamma^{\rm inel}_{p}} \sim \frac{S^{\rm dsa}}{K/L} \,,
\end{equation}
where the spallation rate is neglected for simplicity. The effect of the propagation 
in steepening the spectrum is clearly evident: for a source spectrum $S^{\rm dsa}(E) \propto E^{-\nu}$ and a Galactic 
diffusion coefficient of the type $K(E) \propto E^{\delta}$, the model
predicts that $N(E) \propto E^{-\nu-\delta}$. 
To resolve the two parameters $\nu$ and $\delta$, one has to consider pure secondary species. 
The solution for a one-progenitor secondary CR is given by Eq.\,\ref{Eq::Primary1D},
with the replacement of $S^{\rm dsa}$ with $S^{\rm ism}=\tilde\Gamma^{\rm frag}_{sp}N_{p}$, so that the ratio 
$N_{s}/N_{p} \propto L/K$ allows the simultaneous determination of $\delta$ and $K_{0}/L$. 
These simple trends are valid for the \textit{reference model}, \textit{i.e.} when only primary CRs have 
a source term. In the case of a secondary SNR component, depending on its intensities, 
the parameter determination may be more complicated.  

\section{Analysis and results}  
\label{Sec::Results}            

We now review the model parameters and test the \textit{reference model} setup.
We then analyze the secondary CR production and re-acceleration in
SNRs in some simple scenarios.

\subsection{Model parameters}  
\label{Sec::ModelParameters}   

The DSA mechanism of Sect.\,\S\ref{Sec::SNRAcceleration} provides power-law spectra $S^{\rm dsa} \propto E^{-\nu}$ 
with a unique spectral index $\nu=\alpha-2$ for all the primary CRs, where $\alpha$, in turn, is linked to the 
compression ratio $r$, which is specified by $\delta$ and the observed log-slope $\gamma$.
To match $\gamma\approx 2.7$ with $\delta < 0.7$, one has to adopt a compression factor of $r < 4$, 
in contrast to the value $r=4$ required for strong shocks. Despite this tension between DSA and observations, 
we regard $r$ as an effective quantity describing the compression ratio actually felt by the particles, 
which is not necessary related to the physical strength of the SNR shocks \citep{Ptuskin2010}. 
The diffusion coefficient around the shock is taken to be Bohm-like, $D = \frac{pc}{3 Z B}$. The ambient magnetic 
field $B$ may reach $\sim$\,$100$\,$\mu$G or more, because of amplification effects, except for the very late 
SNR evolutionary stages, where the magnetic field may be damped ($B \lesssim \mu$G).
In our steady-state description, the SNR parameters have to be considered as effective time-averaged quantities
representing a more complex situation where the shock structure evolves with time and may be influenced 
by the back-reaction of accelerated CRs.
The average shock speed $u_{1}$ is of the order of $10^{8}\,$cm\,s$^{-1}$. 
The upstream gas density, $n_{1}$, is poorly known and may well vary from $\sim$\,$10^{-3}$ to $\sim$\,10 \,cm$^{-3}$,
depending on the SNR progenitor star or its local environment. 
The SNR explosion rate per unit volume 
is expressed as a surface density, $2h\mathcal{R}_{\rm snr}$, 
that we fix to 25\,Myr$^{-1}$\,kpc$^{-2}$ \citep{Grenier2000}. 

The parameters describing the interstellar diffusion coefficient are fixed to  $\delta=0.5$ and $K_{0}=$\,0.089\,kpc$^{2}$\,Myr$^{-1}$ (see Sect\,\S\ref{Sec::StandardModelResults}). 
Below the reference rigidity, $R_{0}=$\,4\,GV, we set $\delta=$\,0. 
However our analysis is always applied to rigidities $R>R_{0}$. 
The halo radius is $r_{\rm max}=20\,$kpc and its half-height is $L=5\,$kpc.
As per the propagation in the ISM, the quantity that enters 
the model is surface density $h \times n_{\rm ism}$, 
where we take $h=0.1\,$kpc and $n_{\rm ism} =$\,1\,cm$^{-3}$.
We assume a composition of $90\%\,$\Hyd{}\,$+$\,$10\%\,$\He{} for the ISM gas density, $n_{\rm ism}$, 
and that this composition, on average, is also found in the SNR background media.
In addition, we include the solar modulation effect, though it is
relevant only below a few GeV\,nucleon$^{-1}$. 
The modulation is described in the force-field approximation \citep{Gleeson1968} 
by means of the parameter $\phi$, taken to be $500\,$MV, to characterize a medium-level 
modulation strength. 
Our nuclear chain starts with $Z_{\rm max}=14$ and processes all the relevant isotopes down to $Z=3$.
Nuclei with $Z>14$ do not contribute significantly to the \Li-\Be-\B{} abundances.
The spallation cross-sections are taken from \citet{Silberberg1998}. 
The cross-sections on \He{} targets are obtained by means of the algorithm presented in \citet{Ferrando1988}.
The reference model parameters are listed in Table\,\ref{Tab::Parameters}. 
\setlength{\tabcolsep}{0.036in} 
\begin{table}[!h]
\caption{\captionsize
  Source and transport parameter sets.
\label{Tab::Parameters}}
\centering
\begin{tabular}{ccccc} \hline\hline
\multicolumn{2}{c}{ Acceleration parameters } & {\hspace{1.5cm}} & \multicolumn{2}{c}{Propagation parameters} \\
\hline
$u_{1}$                &  10$^{8}$\,cm\,s$^{-1}$     &     &   $K_0$    & 0.089\,kpc$^{2}$\,Myr$^{-1}$ \\
$B$                &  50.0\,$\mu$G     &     &   $\delta$    &  0.50 \\
$\gamma$                &  2.7     &     &   $R_{0}$    &   4\,GV \\
$n_{1}$                &  1\,cm$^{-3}$     &     &   $L$   &  5\,kpc  \\
$Z_{\rm max}$           &   14     &     &   $r_{\rm max}$    &  20\,kpc \\
$\tau_{\rm snr}$        &   20\,kyr    &     &   $h$    & 0.1\,kpc  \\
$\mathcal{R}_{\rm snr}$  & 125\,Myr$^{-1}$\,kpc$^{-2}$      &     &   $n_{\rm ism}$    &   1\,cm$^{-3}$ \\
$R_{\rm inj}$            &  1\,GV     &     &   $\phi$    &  0.5\,GV \\ 
\hline
\end{tabular}
\end{table}

\subsection{Reference model} 
\label{Sec::StandardModelResults}   
Before analyzing the impact of SNR production and re-acceleration of secondary CRs on the parameter 
determination, we test the \textit{reference model} predictions for the parameters in Table\,\ref{Tab::Parameters}. 
Predictions at Earth for the CR elemental spectra \Li, \Be, \B, \C, \N, \Oxy, \Ne, \Mg, and \Si{}
are presented in Fig.\,\ref{Fig::ccAllSpectra}, where the total spectra (solid lines) are shown
together their secondary component (dashed lines) arising from collisions in the ISM.
The key quantities for propagation, $K_{0}$ and $\delta$, are determined from the \BC{} 
ratio above 2\,GeV\,nucleon$^{-1}$, 
using all the data reported in the past two decades, \textit{i.e.}, from the space-based experiments 
HEAO3-C2 \citep{Engelmann1990}, CRN \citep{Swordy1990}, and \AMS-01 \citep{Aguilar2010}, and 
the balloon-borne projects CREAM \citep{Ahn2009} and ATIC-2 \citep{ATIC2007}.
The primary nuclei spectra were normalized using data from CREAM 
(for \C, \N, \Oxy, \Ne, \Mg, and \Si) and HEAO3-C2 (for all elements).
Given the $K_{0}-L$ degeneracy (see Sect.\,\S\ref{Sec::GalacticPropagation}), 
in the following we adopt the quantity $K_{0}/L$ as 
the physical parameter, where the halo height $L$ is fixed at 5\,kpc.
\\
As apparent from the figure, the \textit{reference model} calculations give a good description 
of the CR elemental spectra within the precision of the present data.
We note that, under this pure diffusion model, 
the B/C data between $\sim$\,10\,GeV and $\sim$\,1\,TeV per nucleon
suggest that $\delta\,\sim\,0.4$, 
while the data at lower energies ($\sim$\,1--100 GeV\,nucleon$^{-1}$) favor higher values ($\delta\,\sim$\,0.6).
These uncertainties are related on both the model unknowns at $\sim$\,GeV/n energies and the  
lack of data at $\gtrsim$\,100\,GeV\,nucleon$^{-1}$ \citep{Maurin2010}.
We also note that the \textit{reference model} predictions are
insensitive to the source parameters $n_{1}$, $u_{1}$, and $B$:
the source spectra are specified only by the effective compression ratio ${r}$ (via $\gamma$ and $\delta$) 
and the abundance constants $Y$. This setup is equivalent to that of many diffusion models, 
\textit{e.g.} \citet{Maurin2001}, that make use of rigidity power-law parameterizations as source functions. 

\subsection{SNR models} 
\label{Sec::SNRModels}  

Similarly to \citet{Morlino2011}, we consider two ideal situations represented by Type I/a 
(important for fragmentation, Sect.\,\S\ref{Sec::SecondaryFragmentFromSNRs}) and 
core-collapse supernovae (important for re-acceleration, Sect.\,\S\ref{Sec::ReAcceleratedNucleiInSNRs}). 
In the Type I/a scenario, the supernova explodes in the regular ISM with typical 
density and temperature $n_{1} \approx 1$\,cm$^{-3}$ and $T_0= 10^4$\,K. 
In the core-collapse scenario, the SNR expands into a hot diluted 
bubble ($n_{1} \ll 1$ cm$^{-3}$ and $T_0 \gg 10^{4}$\,K) that may be generated 
by either the progenitor's wind or by previous SNR explosions that occurred in the same region. 
In both scenarios, the circumstellar densities are assumed to be homogeneous and constant 
during the SNR evolution. Two SNR evolutionary stages are relevant to our study:
the ejecta-dominated (ED) phase, when the shock front expands freely and accumulates 
the swept-up mass in the SNR interior, and the Sedov-Taylor (ST) expansion phase, 
which is driven by the thermal pressure of the hot gas. 
The phase transition ED--ST 
occurs at the time $\tau_{\rm st}$, when the swept-up mass equals the mass of the ejecta $M_{\rm ej}$. 
The CR acceleration ceases at the time $\tau_{\rm snr}$. 

\begin{table}[!h]
\caption{\captionsize
  Case studies of Type I/a and core-collapse SNRs.
\label{Tab::SNRTypes}
}
\centering
\begin{tabular}{ccccc} \hline\hline
\multicolumn{2}{c}{SNR model} & {$n_{1}$ (cm$^{-3}$)} & {$\tau_{\rm st}$ (yr)} & {$\bar{u}_{1}$ (cm\,s$^{-1}$)} \\
\hline
               &  I/a\;\#1  & 0.5    & 330     & 1.3$\times$10$^{8}$  \\
type I/a       &  I/a\;\#2  & 1.5    & 230     & 1.0$\times$10$^{8}$  \\
               &  I/a\;\#3  & 3.0    & 183     & 8.9$\times$10$^{7}$  \\
\hline \\
core           &  CC\;\#1   & 0.003  & 1829    & 3.5$\times$10$^{8}$  \\
collapse       &  CC\;\#2   & 0.01   & 1225    & 2.8$\times$10$^{8}$  \\
               &  CC\;\#3   & 0.1    &  568    & 1.8$\times$10$^{8}$  \\                
\hline
\end{tabular}
\end{table}

For all the models, we assume $E_{\rm snr}= 5\cdot 10^{51}$\,erg, 
$M_{\rm ej}=4 M_{\odot}$, and $\tau_{\rm snr} = 20$\,kyr,
where $E_{\rm snr}$ is the SNR explosion energy (not converted into neutrinos)
and $M_{\odot}$ is the solar mass.
The cases of different $\tau_{\rm snr}$ values are considered in Sect.\,\S\ref{Sec::Summary}.
During the ED stage,
the SNR radius grows with a constant rate $R_{\rm sh}(t) = u_{1} t$, at the speed 
$u_{1} = \left( 2 E_{\rm snr}/M_{\rm ej} \right)^{1/2} \sim$\,10$^{8}$\,cm\,s$^{-1}$,
until it reaches the \textit{swept-up radius} $R_{\rm sw} \equiv R_{\rm sh}(\tau_{\rm st})$. 
For a given SNR model characterized by 
$n_{1}$ and $\tau_{\rm snr}$, we parametrize the shock evolution ($R_{\rm sh}$ and $u_{1}$) using the 
self-similar solutions derived in \citet{Truelove1999} for a remnant expanding into a homogeneous medium. 
These solutions connect smoothly the ED phase ($R_{\rm sh} \propto t$) with the ST stage ($R_{\rm sh} \propto t^{2/5}$). 
The CR acceleration ceases at time $\tau_{\rm snr}$, 
from which we compute the average velocity $\bar{u_{1}} \equiv  R_{\rm sh}(\tau_{\rm snr})/\tau_{\rm snr}$. 
Thus, we use $\bar{u_{1}}$ as an input parameter for our steady-state DSA calculations (Sect.\,\S\ref{Sec::AccelerationDSA}) 
to compute the spectra for all the CR elements. 
The SNR models considered are listed in Table\,\ref{Tab::SNRTypes}.   

We always assume that the total CR flux is produced by SNRs of only one type: 
for each SNR model, the parameters employed have to be regarded as effective ones representing the 
average population of CR sources. We note that this simplified breakdown is somewhat artificial, 
because the total CR flux may be due to a complex ensemble of contributing SNRs.

\subsection{Secondary CR production in Type I/a SNRs} 
\label{Sec::SecondaryFragmentFromSNRs}                

The secondary production of CRs is relevant for SNRs that expand into ambient densities 
of the order of $n_{1}\sim$\,1\,cm$^{-3}$, where the quantity $n_{1}$ represents the average SNR background density. 
Such a value may be 
higher than that of the average ISM, owing to, \textit{e.g.}, contributions from SNRs located in high 
density regions of the Galactic bulge, inside the dense cores of molecular clouds or SNRs expanding 
into the winds of their progenitors. 
\begin{figure}[!ht]
\begin{center}
\includegraphics[width=0.81\columnwidth]{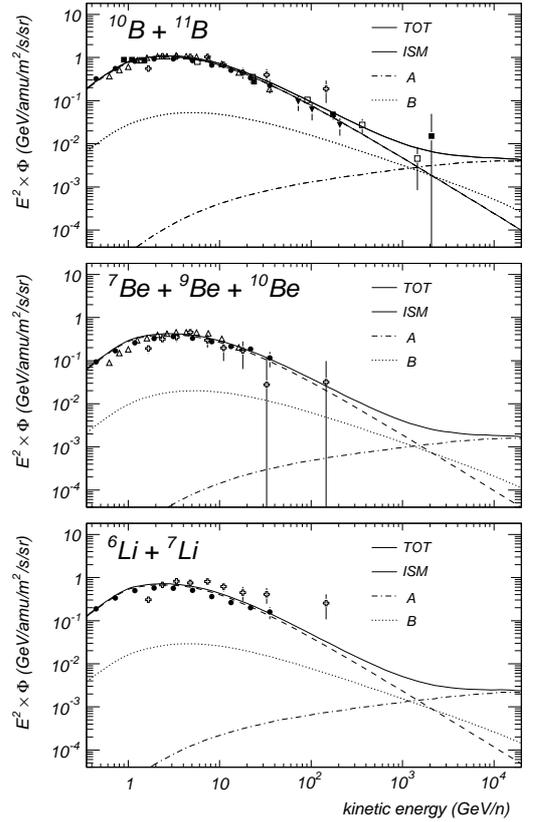}
\caption{\captionsize 
  Energy spectra of secondary elements \Li, \Be, and \B. The source
  components from fragmentation occurring inside SNRs are split into the 
  $\mathcal{A}$-term (dot-dashed lines) and $\mathcal{B}$-term (dotted lines). 
  The dashed lines indicate the ISM-induced components. The solid lines represent the  total spectra. 
  Source parameters are reported in the text. The propagation parameters are as in Table\,\ref{Tab::Parameters}.
  Data as in Fig.\,\ref{Fig::ccAllSpectra}. 
  \label{Fig::ccSecondarySpectra}
}
\end{center}
\end{figure}
\begin{figure}[!h]
\begin{center}
\includegraphics[width=0.81\columnwidth]{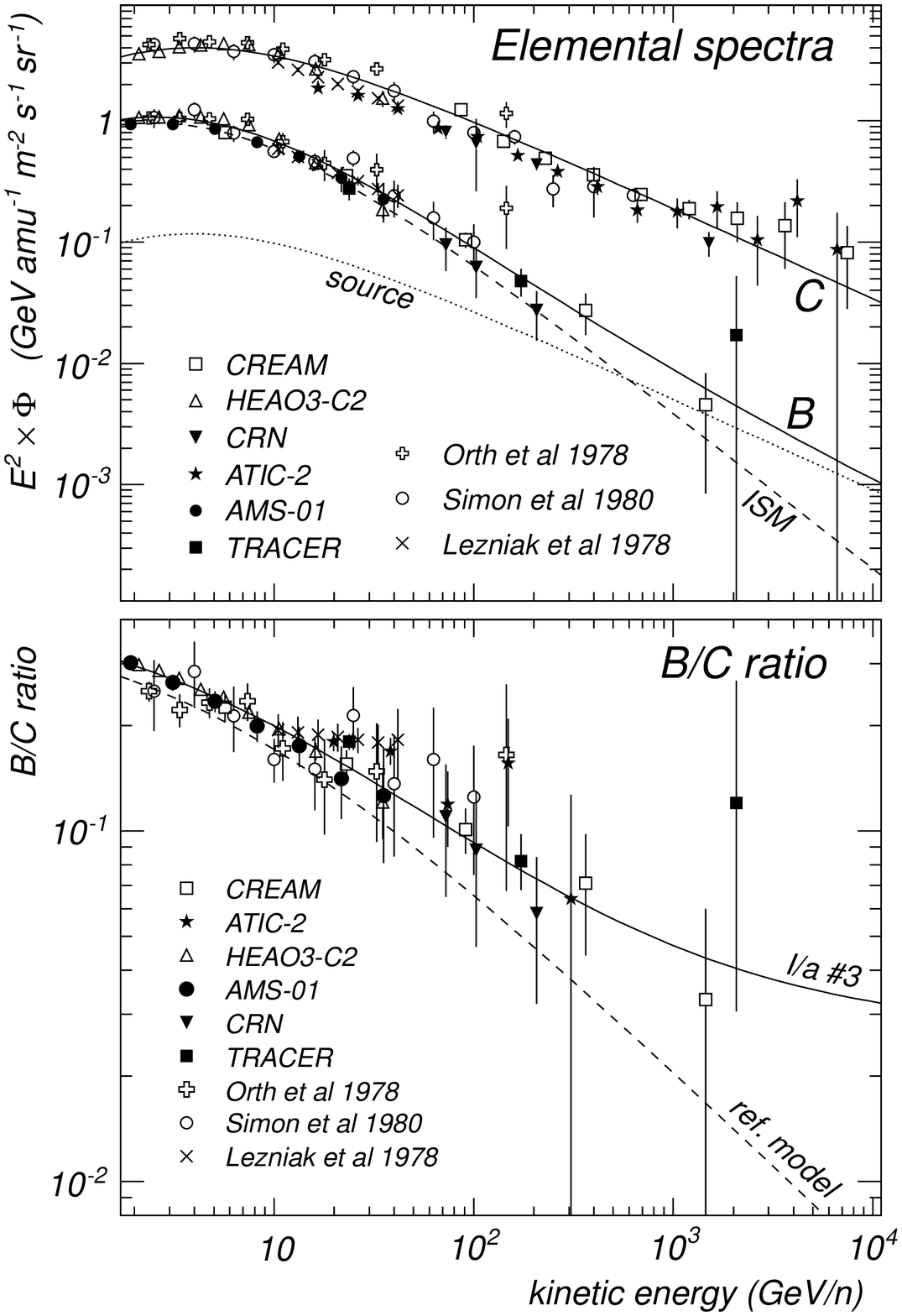}
\caption{\captionsize 
  Top: individual CR spectra of \B{} and \C. 
  Solid lines are the model predictions for I/a\,\#3 SNR model of Table\,\ref{Tab::SNRTypes}.
  Model parameters are as in Table\,\ref{Tab::Parameters}, except for
  $\delta$ and $K_{0}/L$, which are fitted to data.
  The boron SNR component (dotted line) and the ISM component (dashed lines) are reported.
  Bottom: the \BC{} ratio from the above model (solid line) and when fragmentation in SNR is turned off (dashed line).
  Data are from HEAO3-C2 \citep{Engelmann1990}, CREAM \citep{Ahn2009}, \AMS-01 \citep{Aguilar2010},
  TRACER \citep{Obermeier2011}, ATIC-2 \citep{ATIC2007}, CRN~\citep{Muller1991}, \citet{Simon1980}, 
  \citet{Lezniak1978}, and \citet{Orth1978}. 
  \label{Fig::ccBCBestFitFRAG}
}
\end{center}
\end{figure}

\begin{figure*}[!htb]
\begin{center}
\includegraphics[width=1.75\columnwidth]{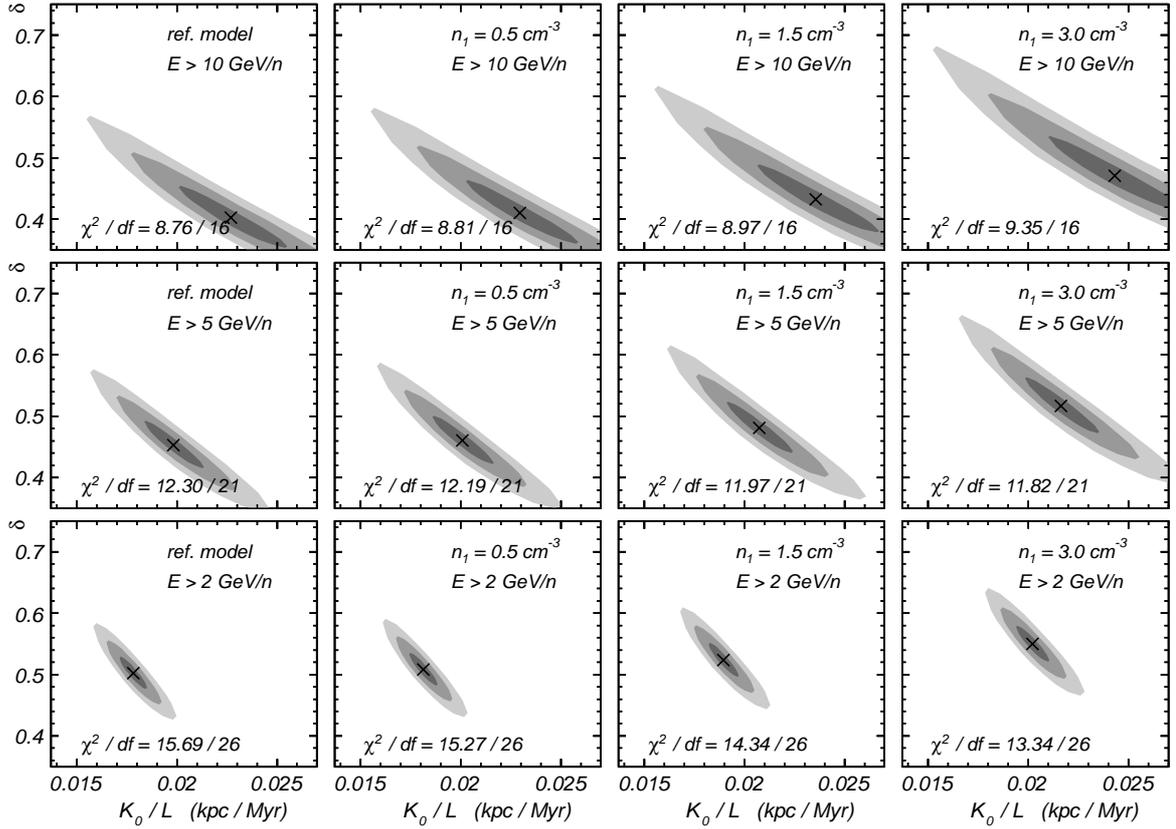}
\caption{\captionsize 
  Fit results for the parameters $\delta$ and $K_{0}/L$ of our models 
  with fragmentation in type I/a SNRs. 
  Results are shown for $E_{\rm min}=$\,2, 5 and
  10\,GeV\,nucleon$^{-1}$ (top to bottom), 
  for the \textit{reference model} and for the SNR models I/a\,\#\,1, \#\,2 and \#\,3 (left to right) of Table\,\ref{Tab::SNRTypes}. 
  The shaded areas represent the 1-, 2- and 3-$\sigma$ contour limits.
  The markers ``$\times$'' indicate the best-fit parameters for each configuration; the $\chi^{2}/df$ 
  ratio is reported in each panel. 
  \label{Fig::ccDeltaVSDiffCoeffFRAG}
}
\end{center}
\end{figure*}
\begin{figure}[!htb]
\begin{center}
\includegraphics[width=0.81\columnwidth]{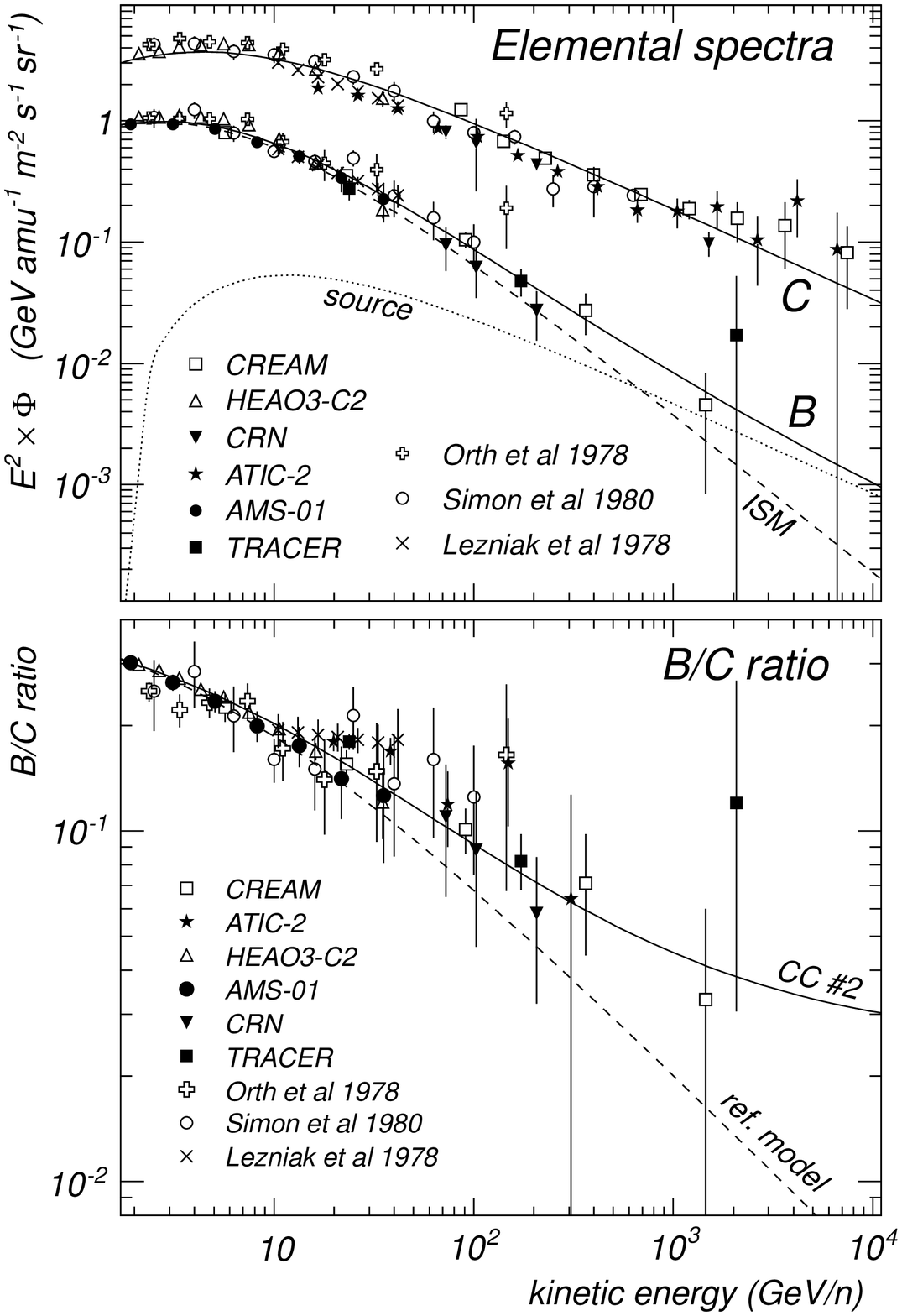}
\caption{\captionsize 
  Top: individual CR spectra of \B{} and \C. 
  Solid lines are the model predictions for CC\,\#2 SNR model of Table\,\ref{Tab::SNRTypes}.
  Model parameters are as in Table\,\ref{Tab::Parameters} except for
  $\delta$ and $K_{0}/L$, which are fitted to data.
  The boron SNR component (dotted line) and the ISM component (dashed lines) are reported.
  Bottom: the \BC{} ratio from the above model (solid line) and when re-acceleration is turned off (dashed line).
  Data as in Fig.\,\ref{Fig::ccBCBestFitFRAG}.   \label{Fig::ccBCBestFitREAC}
}
\end{center}
\end{figure}
From Eq.\,\ref{Eq::DownStreamSecondary}, one sees that the secondary CR flux emitted by 
SNRs has two components. By analogy with \citet{BlasiSerpico2009} and \citet{Kachelriess2011}, these 
are referred to as $\mathcal{A}$ and $\mathcal{B}$. The $\mathcal{A}$--term, proportional to $f_{0}$, 
describes the particles that are produced within a distance $\sim D/u$
of both the sides of the 
shock front and are still able to undergo DSA. From Eq.\,\ref{Eq::GTermSecondary}, the 
$\mathcal{A}$--spectrum is $f \sim p^{-\alpha+1}$, reflecting the spectrum of their progenitors 
($f \sim p^{-\alpha}$) and the momentum dependence of the diffusion coefficient, $D \propto p$.
The $\mathcal{B}$--term, $f \sim q_{2}$, describes those secondary nuclei that, after being 
produced, are simply advected downstream without experiencing further acceleration. 
Their spectrum maintains the same behavior as that of their progenitors $q_{2} \sim p^{-\alpha}$. 
These components are illustrated in Fig.\,\ref{Fig::ccSecondarySpectra} 
for the spectra at Earth of \Li, \Be, and \B{} for a SNR with $n_{1}=2$\,cm$^{-3}$, 
$u_{1}=$\,5\,$\cdot$10$^{7}$\,cm\,s$^{-1}$, and $B=0.1$\,$\mu$G. 
The figure compares the standard ISM components (solid lines) with the source components 
$\mathcal{A}$ (short-dashed lines) and $\mathcal{B}$ (long-dashed lines) within the same 
propagation parameter set. 
Both the source components are harder than those expected from the standard CR production in ISM; 
in particular, the $\mathcal{A}$--term leads to increasing secondary-to-primary ratios at high energies \citep{Blasi2009, Mertsch2009}. 
While the $\mathcal{B}$--term depends on the SNR ambient density $n_{1}$ and its age $\tau_{\rm snr}$, 
the $\mathcal{A}$--term also relies on the diffusion properties, as its strength is proportional to
$\sim \Gamma^{\rm frag}_{kj}D/u_{1}^{2}$.
However, the parameter combination $ n_{1}/(B u_{1}^{2})$ should be
sufficiently large to ensure an $\mathcal{A}$--term dominance 
at $\sim$\,TeV energies, which can be realized only in the latest evolutionary stage of a SNR 
characterized by damped magnetic fields ($B \ll$\,1\,$\mu$G) and low shock speeds ($u_{1}<$\,10$^{8}$\,cm\,s$^{-1}$).
On the other hand, the local flux of stable CR nuclei depends on the
large-scale structure of the galaxy (of some kpc) 
and reflects the contribution of a relatively large population of SNRs and their histories \citep{Taillet2003}.
Furthermore, from Eq.\,\ref{Eq::FullSolutionAtShockFront} and Eq.\,\ref{Eq::ChiApprox}, 
the $\mathcal{A}$--term induces an exponential cut-off at momentum $p^{\rm cut}$ given by $\chi(p^{\rm cut})\approx 1$,
which is not observed in present data of primary or secondary CR spectra.
Since our aim is to estimate these effects when the considered SNRs
produce all the observed CR flux, the associated parameters have to be able to accelerate all CR nuclei 
up to, say, $p^{\rm max}/Z \sim$\,10$^{6}$\,GV.
Thus, from the requirement that $\chi \lesssim 1$ 
for any $p$ up to $p^{\rm max}$
(see Sect.\,\S\ref{Sec::AccelerationDSA}), 
the $\mathcal{A}$-term is always ineffective at the energies we consider 
and is not discussed below. 
Given the absence of a clear spectral feature in the $\mathcal{B}$--term, the spectral deformation 
induced by interactions in SNRs may be difficult to detected $\sim$\,TeV energies, because
it can be easily mimicked by a different choice of $\delta$ and $K_{0}/L$.
%
This is illustrated in Fig.\,\ref{Fig::ccBCBestFitFRAG}, where the total boron spectrum (solid line) is 
plotted showing its standard component arising from ISM collisions (dashed line) and the  
source component coming from hadronic interactions in SNRs (dotted line). The SNR model is the 
I/a\,\#3 of Table\,\ref{Tab::SNRTypes}. We note that the carbon flux also contains a 
small amount of secondary fragments ($\lesssim$\,5\%), produced in both ISM and SNRs. 
The \BC{} ratio is also plotted for the \textit{reference model} (dashed line) under the same propagation 
parameter setting, \textit{i.e.} when hadronic interactions in SNRs are turned off. 
The effect of including secondary production in the sources translates 
into a slight increase at 100 GeV\, nucleon$^{-1}$, while it reaches a factor 
2.5 at 1 TeV\, nucleon$^{-1}$ and 
one order of magnitude at 10 TeV\, nucleon$^{-1}$. 

From the \BC{} ratio data, we determined the parameters $K_{0}/L$ and $\delta$ for the Type I/a SNR 
models of Table\,\ref{Tab::SNRTypes}. 
We performed a $\chi^{2}$ analysis using our model interfaced with \texttt{MINUIT}. 
The data were fit above a minimal energy of $E_{\rm min}=$\,10\,GeV\,nucleon$^{-1}$, 
as a compromise between the diffusion--dominated regime and the availability of experimental data. 
We also repeated the fits down to lower $E_{\rm min}$ to test the relevance of 
low energy effects to our model. The results are shown in Fig.\,\ref{Fig::ccDeltaVSDiffCoeffFRAG} 
for $E_{\rm min}=$\,2, 5, and 10\,GeV\,nucleon$^{-1}$ (from top to
bottom), for \textit{reference model} and I/a models of
Table\,\ref{Tab::SNRTypes} (left to right). 
The shaded areas represent the 1--, 2--, and 3--$\sigma$ contour limits of the $\chi^{2}$. 
We stress that these parameter uncertainties are those arising from
the fits and that they are contextual to our models. 
Owing to the complexity of the physics processes involved together with the possible 
lack of knowledge of several astrophysical inputs, the actual parameter uncertainties may be 
much larger \citep{Maurin2010}. For instance, the  published values of $\delta$ vary widely 
from $\sim$\,0.3 to $\sim$\,0.7.
The markers describe the best-fit parameters for each configuration. The $\chi^{2}$ values reported
in each panels are divided by the degrees of freedom $df=$\,26, 21, and 16 for the considered energy 
thresholds. It can be seen from Fig.\,\ref{Fig::ccDeltaVSDiffCoeffFRAG} 
that the source component has a little effect for model 
I/a\,\#3 ($n_{1}=$\,0.5\,cm$^{-3}$). 
When denser media were considered, the secondary source component was
found to flatten the \BC{} ratio, so that
higher values of $\delta$ were required to match the data. This trend is clearly apparent in 
Fig.\,\ref{Fig::ccDeltaVSDiffCoeffFRAG} (from left to right). Similar conclusions, though weaker, 
can be drawn for the $K_{0}/L$ parameter ratio. To first approximation, \BC\,$\propto$\,$L/K_{0}$, so that 
the presence of a SNR component of boron requires a larger $K_{0}/L$ ratio to match the data. 

In summary, for the SNR models considered, the fragmentation in SNRs affects the parameter $\delta$ of $\sim$\,5--15\% (and $K_{0}/L$ of $\sim$\,2--10\%), but these models cannot be discriminated 
by present data because of the large uncertainties in the data.
This $n_{1}$--$\delta$ degeneracy is apparent by the $\chi^{2}/df$--values, which are 
almost insensitive to the SNR properties.

\subsection{Re-acceleration in core-collapse SNRs}  
\label{Sec::ReAcceleratedNucleiInSNRs}              

The amount of re-accelerated CRs depends on the total volume occupied by the SNRs 
(per unit time) and their explosion rate (per unit volume). The fraction of re-accelerated CRs 
to the total background CRs can be roughly estimated as $N^{\rm re}/N^{\rm bg} \sim V_{\rm snr}\mathcal{R}_{\rm snr}\tau_{\rm esc}$, 
where $V_{\rm snr}$ is the SNR volume and $\tau_{\rm esc}$ is the characteristic escape time of CRs in 
the Galaxy. At a few GeV\,nucleon$^{-1}$, $\tau_{\rm esc} \sim 2hL/K \sim$\,5\,Myr. 
The $V_{\rm snr}$ is mainly determined by its expansion during the ED phase;
the SNR reaches a spherical volume 
$V_{\rm sw} =  { M_{\rm ej}}/{\left( \bar{m} n_{1}\right)}$,
where $\bar{m}$ is the mean mass of the ambient gas. 
Thus $N^{\rm re}/N^{\rm bg} \propto  1/n_{1}$, which is an opposite
trend to that of the fragmentation 
scenario of Sect.\,\S\ref{Sec::SecondaryFragmentFromSNRs}. 
One can see that for a density $n_{1}\sim$\,1\,cm$^{-3}$ the re-acceleration gives a small contribution to the total CR flux. 
In contrast, for $n_{1}\lesssim\,$0.01\,cm$^{-3}$, the re-acceleration fraction grows significantly ($\gtrsim$ few percent).
However it is also important the subsequent ST phase, where the SNR shock expands adiabatically as 
$R_{\rm sh}(t) \propto t^{2/5}$, slowing down at the rate $u_{1}(t) = t^{-3/5}$.

Using the SNR parameters of Table\,\ref{Tab::SNRTypes},
we computed the re-accelerated CR spectra as in Sect.\,\S\ref{Sec::PrimaryNuclei}, 
using  $Q^{\rm reac} =  f^{\rm bg}(p) \delta(x)$, where $f^{\rm bg}(p) = \frac{\beta N(p)}{4\pi A p^{2}}$. 
Since CRs are already supra-thermal, 
we assumed that all CR particles above $p^{\rm inj}$ are suitable for (re-)undergoing DSA. 
We note that $N(p)$ is the DHM solution of Eq.\,\ref{Eq::DHMEquation} that, in turn, is fed by the total DSA spectra. 
Hence, we  solved the DSA and DHM equation systems iteratively. 
At the first iteration, only the standard injection term was considered 
(Eq.\,\ref{Eq::Injection}) to compute the interstellar flux $N$ for all  
CR nuclei. The subsequent iterations made use of the previous DHM solutions, $N$, 
to update the terms $Q^{\rm pri}$ and $Q^{\rm reac}$ and to re-compute the total interstellar fluxes. 
The procedure was iterated until the convergence was reached. 
At each iteration, the injection constants, $Y$, were re-adjusted. The resulting CR flux (standard plus re-accelerated) 
was therefore determined by Eq.\,\ref{Eq::SNRVolumeIntegral} and is fully specified by the source parameters 
$n_{1}$ and $\tau_{\rm snr}$. In practice, we found that five iterations ensure a stable solution.  
The effect of re-acceleration is shown in Fig.\,\ref{Fig::ccBCBestFitREAC} for the SNR model  
CC\,\#2 of Table\,\ref{Tab::SNRTypes}. At energies of $\sim$\,1\,TeV\,nucleon$^{-1}$, the re-accelerated 
component dominates over the ISM-induced component for secondary nuclei. 
\begin{figure*}[!ht]
\begin{center}
\includegraphics[width=1.75\columnwidth]{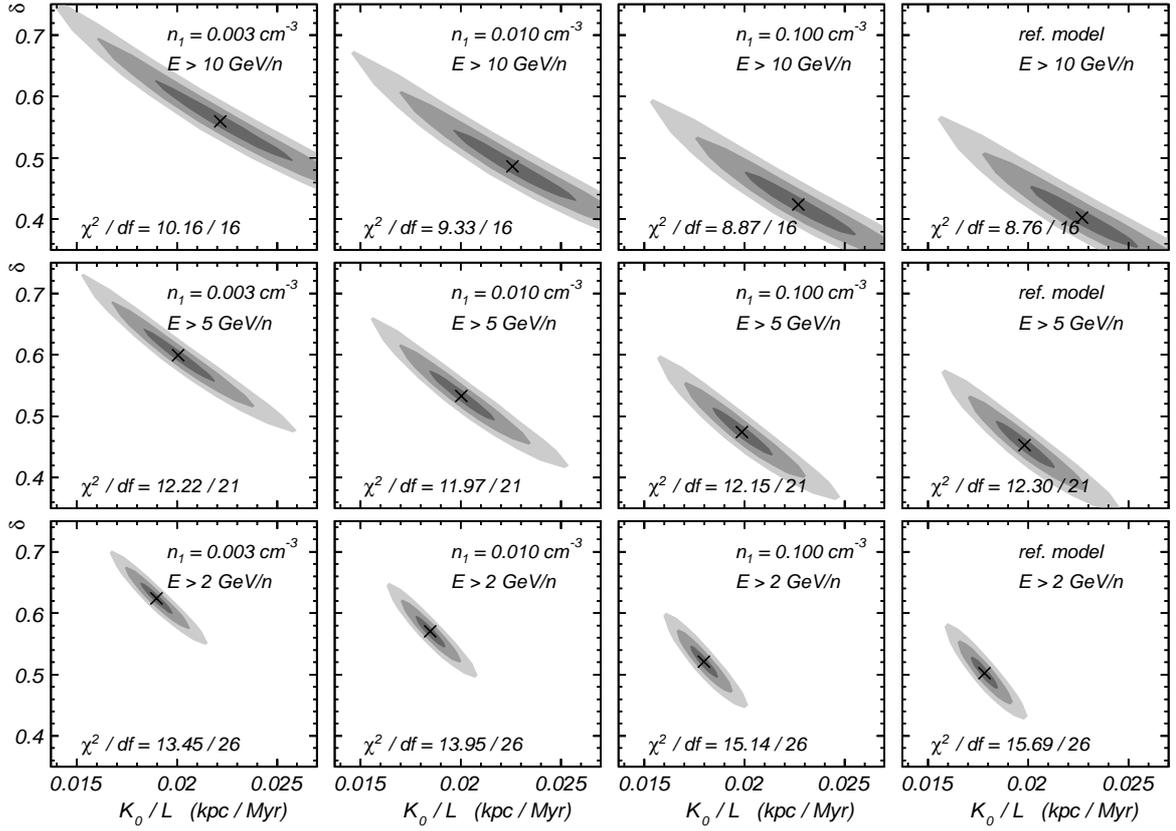}
\caption{\captionsize 
  Fit results for the parameters $\delta$ and $K_{0}/L$ of our models 
  with re-acceleration in core-collapse SNRs. 
  Results are shown for $E_{\rm min}=$\,2, 5, and
  10\,GeV\,nucleon$^{-1}$ (top to bottom), 
  for the SNR models CC\,\#\,1, \#\,2, and \#\,3 
  of Table\,\ref{Tab::SNRTypes} and for the \textit{reference model} (left to right).
  The shaded areas represent the 1-, 2-, and 3-$\sigma$ contour limits. 
  The markers ``$\times$'' indicate the best-fit parameters for each configuration; the $\chi^{2}/df$ 
  ratio is reported in each panel.
  \label{Fig::ccDeltaVSDiffCoeffREAC}
}
\end{center}
\end{figure*}
It should be noted that the sources of re-accelerated CRs may have a complex spatial 
distribution depending on the SNR spatial profile. 
In our model, we used a uniform distribution, $s(r)\equiv$\,1, which 
does not limit the predicting power of diffusion models as long as the key parameters are 
regarded as effective quantities tuned to agree with the data \citep{Maurin2001}. 
However, further elaborations would require more refined descriptions. 
In Fig.\,\ref{Fig::ccDeltaVSDiffCoeffREAC}, we plot the fit results
for the parameters $\delta$ and $K_{0}/L$ for the core-collapse SNR models of
Table\,\ref{Tab::SNRTypes} and the \textit{reference model}.
Compared to the scenario of Sect.\,\S\ref{Sec::SecondaryFragmentFromSNRs}, 
the results are less trivial to interpret because the background CR flux that is subjected to re-acceleration 
itself depends on the parameters $\delta$ and $K_{0}/L$. 
As consequence of this non-linearity, the $K_{0}/L$ best-fit values
results are less sensitive to $n_{1}$.
The results for $\delta$  are qualitatively similar to those of Sect.\,\S\ref{Sec::SecondaryFragmentFromSNRs},
showing an opposite dependence on $n_{1}$. For the SNR model CC\,\#\,1 ($n_{1}=0.003$\,cm$^{-3}$, left column), 
the source component dominates the secondary CR flux 
at $\sim$\,100\,GeV\,nucleon$^{-1}$, so that, at higher energies, the \BC{} ratio becomes appreciably flat.
The effect becomes less significant for higher background densities, \textit{e.g.}, 
CC\,\#\,1 ($n_{1}=0.1$\,cm$^{-3}$, right column) where the best-fit parameters are close 
to those arising from the \textit{reference model} fit.
As for the scenario of Sect.\,\S\ref{Sec::SecondaryFragmentFromSNRs}, results are limited 
by the sizable uncertainties in the parameters that preclude quantitative conclusions  
for $E_{\rm min} = 10$\,GeV\,nucleon$^{-1}$. Nonetheless, the figure shows clear trends, 
especially for $\delta$. The $\chi^{2}/df$ values reported in each panel indicates 
that good fits can be made for all the considered configurations, though they do not 
vary significantly among the various SNR models.

\subsection{Summary and discussion}  
\label{Sec::Summary}                 

Our breakdown into Type I/a and core-collapse SNR scenarios is motivated by the complementary dependence of the two effects on $n_{1}$.
As seen in Sect.\,\S\ref{Sec::SecondaryFragmentFromSNRs} and Sect.\,\S\ref{Sec::ReAcceleratedNucleiInSNRs},
the CR re-acceleration is found to be important for SNRs exploding into rarefied media, which are 
typical of super-bubbles (including our own \textit{local bubble}), 
while the secondary production in SNRs is relevant for ambient densities similar to those of 
the regular ISM. 
Our calculations of the \BC{} ratio are in substantial agreement with the work of 
\citet{Berezhko2003} in the cases where the comparison can be made,
though those authors used different 
approaches to model the acceleration as well as the interstellar propagation. 
The fit results for the SNR models of Table\,\ref{Tab::SNRTypes} 
and the \textit{reference model} are listed in Table\,\ref{Tab::SummaryTable}.

\begin{table*}[!ht]
\caption{\captionsize
  Summary of the fit results to the propagation parameters. See the text in Sect.\,\S\ref{Sec::Summary}.
  \label{Tab::SummaryTable}
}
\centering
\begin{tabular}{c c ccc c ccc c ccc} \hline\hline
 {Fit} & {} &  \multicolumn{3}{c}{ $E>$\,2 GeV/n } &   {} & \multicolumn{3}{c}{ $E>$\,5 GeV/n } &   {} & \multicolumn{3}{c}{ $E>$\,10 GeV/n }\\
 \cline{1-1}\cline{3-5}\cline{7-9}\cline{11-13}\\
  {Model}  & {} & {$\delta$} &  {$K_{0}/L$\,(kpc/Myr)} & {$\chi^{2}/df$} & {} & {$\delta$} &  {$K_{0}/L$\,(kpc/Myr)} & {$\chi^{2}/df$} & {} & {$\delta$} &  {$K_{0}/L$\,(kpc/Myr)} & {$\chi^{2}/df$}\\
\hline
Ref. Model & & 0.50\,$\pm$\,0.03 & 0.01781\,$\pm$\,0.00069  & 15.69/26 & &   0.45\,$\pm$\,0.04   & 0.01982\,$\pm$\,0.00149 & 12.30/21  &  & 0.40\,$\pm$\,0.05  & 0.02270\,$\pm$\,0.00270 & 8.76/16  \\  
SNR I/a \#\,1     & & 0.51\,$\pm$\,0.03  & 0.01813\,$\pm$\,0.00071  & 15.27/26  & &  0.46\,$\pm$\,0.04  & 0.02008\,$\pm$\,0.00153  & 12.19/21  & &  0.41\,$\pm$\,0.05  & 0.02296\,$\pm$\,0.00280  & 8.81/16   \\  
SNR I/a \#\,2     & & 0.52\,$\pm$\,0.03  & 0.01895\,$\pm$\,0.00075  & 14.34/26  & &  0.48\,$\pm$\,0.04  & 0.02073\,$\pm$\,0.00165  & 11.97/21  & &  0.43\,$\pm$\,0.06  & 0.02357\,$\pm$\,0.00304  & 8.97/16   \\  
SNR I/a \#\,3     & & 0.55\,$\pm$\,0.03  & 0.02022\,$\pm$\,0.00083  & 13.34/26  & &  0.52\,$\pm$\,0.05  & 0.02164\,$\pm$\,0.00185  & 11.82/21  & &  0.47\,$\pm$\,0.06  & 0.02432\,$\pm$\,0.00346  & 9.35/16   \\  
SNR CC \#\,1     & & 0.62\,$\pm$\,0.03  & 0.01897\,$\pm$\,0.00082  & 13.45/26  & &  0.60\,$\pm$\,0.04  & 0.02005\,$\pm$\,0.00179  & 12.22/21  & &  0.56\,$\pm$\,0.07  & 0.02217\,$\pm$\,0.00343  & 10.16/16   \\  
SNR CC \#\,2     & & 0.57\,$\pm$\,0.03  & 0.01847\,$\pm$\,0.00076  & 13.95/26  & &  0.53\,$\pm$\,0.04  & 0.02003\,$\pm$\,0.00165  & 11.97/21  & &  0.49\,$\pm$\,0.06  & 0.02258\,$\pm$\,0.00308  & 9.33/16   \\  
SNR CC \#\,3     & & 0.52\,$\pm$\,0.03  & 0.01798\,$\pm$\,0.00070  & 15.14/26  & &  0.47\,$\pm$\,0.04  & 0.01988\,$\pm$\,0.00152  & 12.15/21  & &  0.42\,$\pm$\,0.05  & 0.02269\,$\pm$\,0.00279  & 8.87/16   \\ 
\hline
\end{tabular}
\end{table*}

\begin{figure*}[!ht]
\begin{center}
\includegraphics[width=1.70\columnwidth]{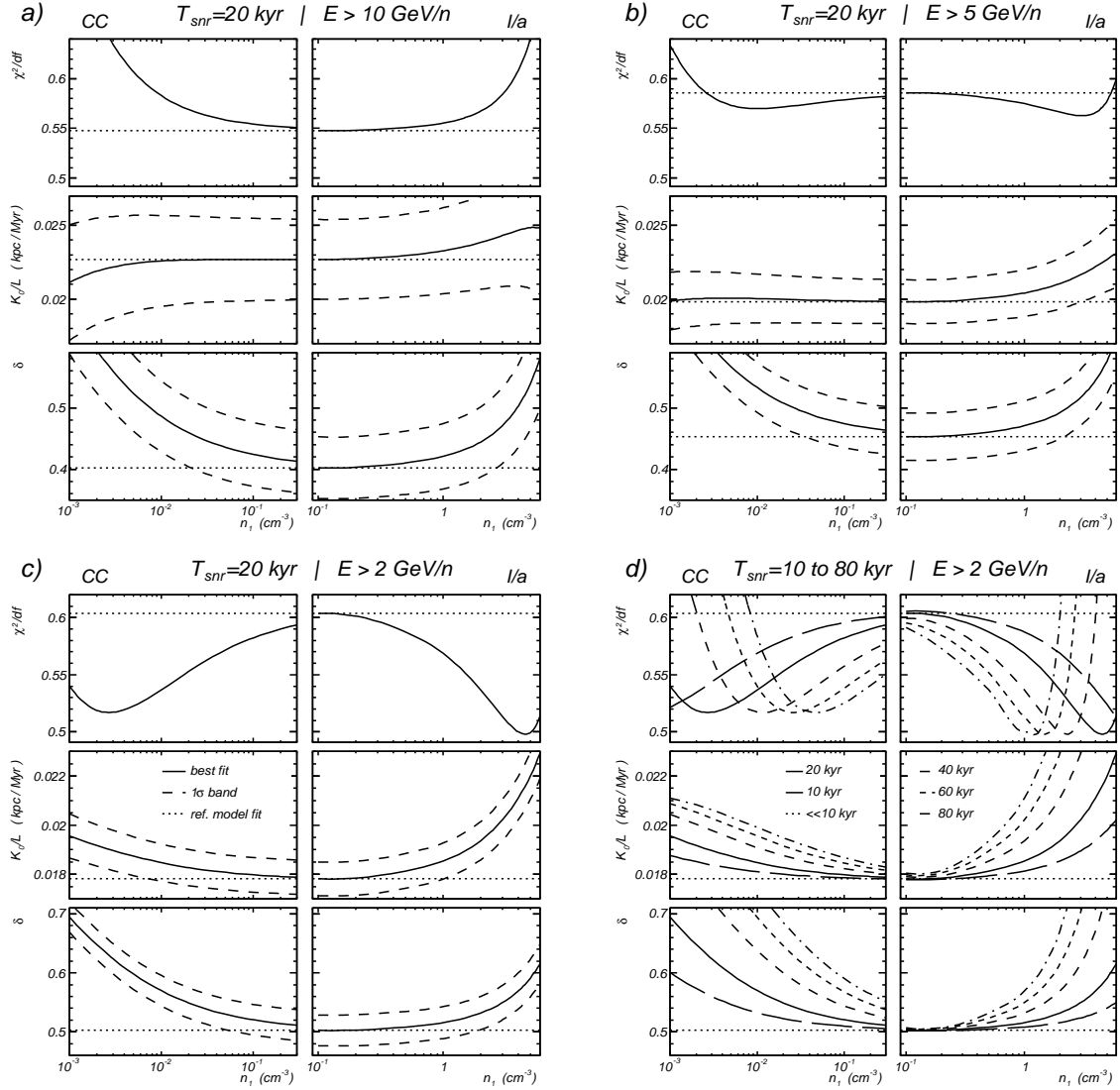}
\caption{\captionsize 
  Best-fit parameters $\delta$ and $K_{0}/L$ and the corresponding $\chi^{2}/df$ as function of $n_{1}$ (solid lines) 
  for models with re-acceleration in core-collapse SNRs (\textit{CC}, left sub-panels) 
  and with hadronic interactions in Type I/a SNRs (\textit{I/a}, right sub-panels).
  The panel groups (a), (b) and (c) show the fit results for data at $E_{\rm min}=$\,1, 10, 5, and 2\,GeV\,nucleon$^{-1}$, respectively,
  for $\tau_{\rm snr}=$\,20\,kyr. Panel (d) shows the results for $\tau_{snr}=$\,10, 20, 40, 60, and 80\,kyr 
  in the case of $E_{\rm min}=$\,2\,GeV\,nucleon$^{-1}$. 
  The horizontal dotted lines indicate the \textit{reference model} parameters.
  \label{Fig::ccFitResultsVSDensity} 
}
\end{center}
\end{figure*}

Figure\,\ref{Fig::ccFitResultsVSDensity} summarizes our findings, showing the 
 best-fit parameters as functions of the SNR circumstellar density.
The panel groups (a), (b), and (c) are referred to fits performed at different minimal 
energies, $E_{\rm min}=$\,10, 5, and 2\,GeV\,nucleon$^{-1}$, respectively. For each group, 
we report $\delta$, $K_{0}/L$, and $\chi^{2}/df$ as functions of $n_{1}$ (from bottom to top, 
solid lines). The two mechanisms are presented separately: the sub-panels on the 
left-hand side  show the effect of re-acceleration in CC type SNRs, while the right-hand side plots 
are referred to the secondary production by spallations in Type I/a SNRs. The complementarity 
of the two effects is apparent from the figure. In the region where they overlap, 
$n_{1}\approx$\,0.5\,cm$^{-3}$, neither is relevant. 
The horizontal (dotted) lines indicate the best-fit parameters for the \textit{reference model}. 
Their dependence on $E_{\rm min}$ resembles that found in \citet{DiBernardo2010}, 
who also considered diffusive reacceleration models, 
although we note that we used a different set of data for the parameter determination.
As discussed, the \textit{reference model} is insensitive to either $n_{1}$ or other SNR parameters. 
The dashed lines indicate the parameter uncertainties (at one $\sigma$ of CL) arising 
from the fits. 

It is interesting to note the evolution of the best-$\chi^{2}$ structures when the minimal 
energy $E_{\rm min}$ is decreased from 10\,GeV\,nucleon$^{-1}$ (Fig. \,\ref{Fig::ccFitResultsVSDensity}a) 
to 2\,GeV\,nucleon$^{-1}$ (Fig.\,\ref{Fig::ccFitResultsVSDensity}c).
When the low-energy \BC{} data are included in the fits, the $\chi^{2}/df$ distribution exhibits 
two minima. The \BC{} ratio data at low energies favor a slope ($\delta\,\sim$\,0.6),
which is somewhat steeper than that observed in high energy data ($\delta\,\sim$\,0.4):
these two regimes are matched by models of SNRs that emit secondary nuclei.
In some previous studies, \textit{e.g.} \citet{Trotta2011}, 
it has been found that 
the secondary-to-primary ratios can be reproduced well at all energies using $\delta=1/3$ and a strong 
diffusive reacceleration (the interstellar Alfv\'enic speed is of the order of $\sim$\,30\,km\,s$^{-1}$). 
However, this cannot be satisfactorily reconciled with the use of pure power-law functions for the CR sources.
On the other hand, the trends we observe suggest a possible role of SNR-fragmentation or re-acceleration 
in reconciling the low-energy \BC{} data with those at higher energies in a pure diffusion scenario.

As we have stressed, the physical effects discussed in this work should be tested at high energies, 
where most of the complexity of the low-energy CR propagation can be neglected. 
Owing to the scarcity of CR data above 10\,GeV\,nucleon$^{-1}$, the parameter constraints reported 
in Fig. \,\ref{Fig::ccFitResultsVSDensity}a do not allow us to make any firm discrimination among the different SNR models.
However, the main trends are apparent. 
On the propagation side, all the non-standard scenarios point toward larger values for $\delta$, 
which is in some tensions with the predictions for the interstellar turbulence \citep{Strong2007}
and with CR anisotropy studies \citep{Ptuskin2006}. 
On the acceleration side, large values of $\delta$ reduce the source spectral index 
closer to the value $\nu=$\,2, which is favored by the DSA theory for strong shocks.

In all these scenarios, the acceleration ceases at $\tau_{\rm snr}=$\,20\,kyr,
which may not be the case given their different SNR evolutionary properties. 
For instance, since the ST phase duration scales as $n_{1}^{-4/7}$ \citep{Truelove1999},
one may expect core-collapse SNRs to have longer $\tau_{\rm snr}$ than Type I/a SNRs.
However, the parameter $\tau_{\rm snr}$ represents the time for which
the SNR is active as a CR factory and it can be extremely difficult to estimate. 
Thus, in Fig.\,\ref{Fig::ccFitResultsVSDensity}d, we give the fit results for different 
values of $\tau_{snr}$ from 10\,kyr to 80\,kyr ($E_{\rm min}=$\,2\,GeV\,nucleon$^{-1}$).
The effect of using different $\tau_{\rm snr}$ is clear. 
The longer the time for which the SNR is active, the larger the fragments produced in its interior. 
In practice, the secondary CRs production in Type I/a SNRs is
characterized by the product $n_{1}\tau_{\rm snr}$. 
For re-acceleration, a longer lifetime allows the SNR to occupy larger volumes. 
To first approximation, 
the intensity of re-accelerated nuclei increases as $\sim \tau_{\rm snr}/n_{1}$. 
As shown in the figure, for longer values of $\tau_{\rm snr}$, the re-acceleration 
effect also becomes important for relatively high density media.

\section{The projected \AMS-02 sensitivity}  
\label{Sec::AMSPhysics}                      

We switch now to some estimations for the \AMS{} experiment\footnote{ \url{http://www.ams02.org}}, 
which is devoted to direct measurements of Galactic CRs across a wide range of energy.
\begin{figure*}[ht!]
\begin{center}
\includegraphics[width=1.75\columnwidth]{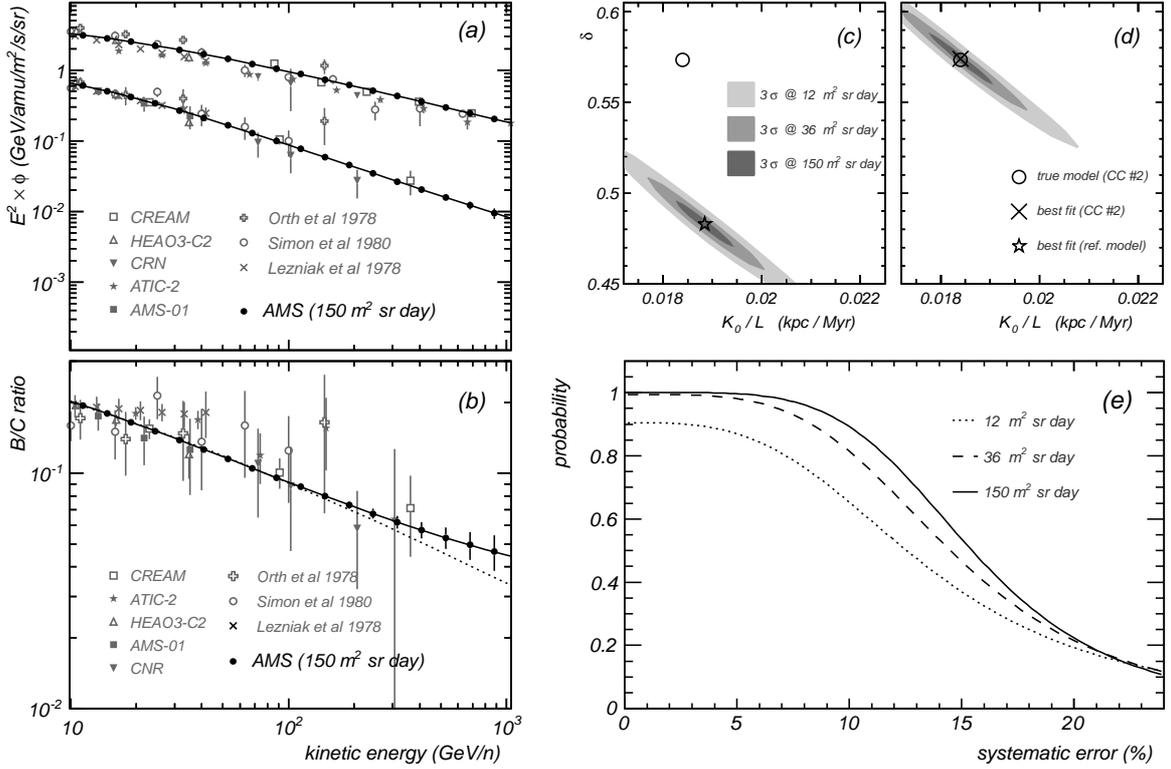}
\caption{\captionsize 
  \AMS-02 mock data for the elemental fluxes $\phi_{B}$ and $\phi_{C}$ (a) and their ratio (b), using the input model
  CC\,\#\,2 of Table\,\ref{Tab::SNRTypes} and assuming a detector exposure factor F= 150\,m$^{2}$\,sr\,day. 
  The error bars are only statistics.
  The constraints to the transport parameters provided by the \AMS-02 mock data are reported by the 3-$\sigma$ contour levels 
  for exposure factors of $F=$\,12, 36, and 150\,m$^{2}$\,sr\,day in (c) and (d). The data are fitted within the reference 
  model (star in panel c and dotted line in panel b) and within the model CC\,\#\,2 
  (cross in panel d and solid line in panel b).
  The \AMS-02 discrimination probability between the two models as a function of the systematic error in the measurement
  is shown in (d) for $F$= 12, 36, and 150\,m$^{2}$\,sr\,day. The systematic errors are assumed to be energy-independent.
  \label{Fig::ccBCBestFitAMSPotential} 
}
\end{center}
\end{figure*}
The prime goals of the \AMS{} project are the direct search for
anti-nuclei and the indirect search for dark matter particles. 
The first version of the experiment, \AMS-01, operated in a test flight on June 1998. 
The final version of experiment, \AMS-02, was successfully installed in the \textit{International Space Station} on May 2011
and will be active for at least ten years. \AMS-02 is able to identify
CR elements from $Z=1$ to $Z=26$ and to determine their energy 
spectra from $\sim$\,0.5\,GeV to $\sim$\,1\,TeV per nucleon with unprecedented accuracy.
We estimate the \AMS-02 capabilities in determining the CR propagation properties for the considered scenarios.

\subsection{Projected data} 
\label{Sec::ProjectedData}  

The \AMS-02 sensitivity to CR nuclei measurements is studied by the generation of \textit{mock} data for a given input model. 
The number of $j$--type particles recorded by \AMS-02 at the kinetic energies between $E_{1}$ and $E_{2}$ is given by
\begin{equation}\label{Eq::DetectorModel}
\Delta N_{j} = \int_{E_{1}}^{E_{2}} \phi_{j}(E)\cdot \mathcal{E}_{j} \cdot \mathcal{G}_{j} \cdot \mathcal{T}_{j}\cdot dE \,,
\end{equation}
where $\phi_{j}$ is the input spectrum, $\mathcal{G}_{j}$ is the detector geometric factor, 
$\mathcal{E}_{j}$ is the detection efficiency, and $\mathcal{T}_{j}$ is the exposure time. 
All of these quantities are in general energy-dependent and
particle-dependent. The relevant quantity for 
our estimates is the exposure factor $\mathcal{F}\equiv \mathcal{E}  \mathcal{G}  \mathcal{T}$, 
which we assume to be both energy and particle independent. 
We consider the cases of $\mathcal{F}=$\,12, 36, and 150\,m$^{2}$\,sr\,day.
For values of, \textit{e.g.}, $\mathcal{G}=$\,0.45\,m$^{2}$\,sr and $\mathcal{E}=$\,90\%,
our choices correspond to one month, three months and one year of time exposures, respectively.
We adopt a log-energy binning using nine bins per decade between 10\,GeV and 1\,TeV per nucleon.
The \AMS-02 mock data for the \B{} and \C{} fluxes and their ratios are shown in 
Fig.\,\ref{Fig::ccBCBestFitAMSPotential}a and \ref{Fig::ccBCBestFitAMSPotential}b. 
The CR fluxes, $\phi_{B}$ and $\phi_{C}$, are calculated using the SNR model 
CC\,\#\,2 (Table\,\ref{Tab::SNRTypes}) as an input model. 
This ``true'' model is characterized by the SNR parameters $n_{1}=$\,0.01\,cm$^{-3}$ and 
$\tau_{\rm snr}=$\,20\,kyr, and transport parameters $K_{0}/L=$\,0.01847\,kpc\,Myr$^{-1}$ 
and $\delta=$\,0.57. From Eq.\,\ref{Eq::DetectorModel}, we compute the statistical 
error in each \BC{} data point as $1/\sqrt{\Delta N_{B}} + 1/\sqrt{\Delta N_{C}}$.
Nonetheless, CR measurements are also affected by  systematic errors, which become 
increasingly important as the precision increases with the collected
statistics.

\subsection{Discrimination power}  
\label{Sec::DiscriminationPower}   

We fit the \BC{} ratio mock data leaving $K_{0}/L$ and $\delta$ as free parameters. These parameters 
are determined within both the \textit{reference model} (re-acceleration off) and the ``true'' re-acceleration 
model CC\,\#\,2 of Table\,\ref{Tab::SNRTypes}. 
As shown in Fig.\,\ref{Fig::ccBCBestFitAMSPotential}b, both the models can be tuned to reproduce 
 the \AMS-02 mock data, but they exhibit different functional shapes and deviate at high energies.
The contour plots in panels (c) and (d) correspond to the best-fit parameters of the two models. 
Contour levels are shown for 3-$\sigma$ uncertainty levels corresponding to the three exposure factors $\mathcal{F}$.
As expected, the re-acceleration model fit (d) returns the
correct parameters, while the \textit{reference model} 
fit (c) misestimates the parameters because of inaccurate assumptions about the source properties.
In fact, when the \textit{reference model} is forced to describe the data, the spectral distortion induced by the 
re-acceleration is mimicked by the use of a lower value for $\delta$. 
Given the precision of the \AMS-02 data, this represents the dominant ``error'' in the parameter determination. 
As apparent from the figure, the \AMS-02 data place tight constraints on the propagation parameters. 
For instance, $\delta$ is determined to a precision better than
$\lesssim$\,10\% within a 3--$\sigma$ uncertainty level.
The $\delta$--$n_{1}$ degeneracy may be lifted as in \citet{Castellina2005}, \textit{i.e.}, 
by a statistical test to discriminate between the two fits. As long as only statistical errors are considered, 
we find that the discrimination between the two scenarios is always possible for the three considered 
exposures at 90\,\% of CL. The effect of systematic errors in the data is shown in 
Fig.\,\ref{Fig::ccBCBestFitAMSPotential}e, where we plot the \AMS-02 discrimination probability versus 
the relative systematic error. Our calculation assumes constant systematic errors (added in quadrature 
to the statistical ones), but these considerations also hold for energy-dependent systematic errors if 
their energy rise is less pronounced than the statistical errors. The solid, dashed, and dotted 
lines represent the cases of $\mathcal{F}=$\,12, 36, and 150\,m$^{2}$\,sr\,day, respectively. To 
achieve a discrimination of 90\% CL, the systematic error has to be smaller than $\sim$\,4\%, 8\%, and 10\% 
for the three considered exposures. A 95\% CL requirement also needs $\mathcal{F}$ to be larger than 
12\,m$^{2}$\,sr\,day. We consider these requirements as reasonable for \AMS-02, because the measurements 
of elemental ratios are only mildly sensitive to systematic errors.
\begin{figure}[!ht]
\begin{center}
\includegraphics[width=0.81\columnwidth]{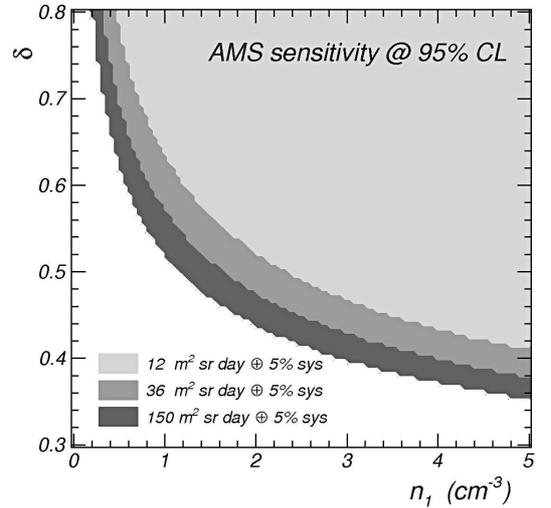}
\caption{\captionsize 
  \AMS-02 discrimination power for models with secondary production in SNRs.
  Each point in the $(n_{1},\delta)$--plane represents an input model with fragmentation inside SNRs
  with $\tau_{\rm snr}=$\,20\,kyr. 
  The parameter $K_{0}/L$ is taken to match the existing \BC{} ratio data. 
  The shaded areas cover the parameter region where \AMS-02 is sensitive at 95\%\,CL 
  for the exposure $\mathcal{F}=$\,12, 36, and 150\,m$^{2}$\,sr\,day. 
  The systematic errors are assumed to be 5\% of the measured \BC{} and constant in energy.
  \label{Fig::ccAMSProjection} 
}
\end{center}
\end{figure}

Similar conclusions can be drawn for models with fragmentation in SNRs. In this case, we have explored 
a large region of the parameter space $n_{1}$--$\delta$, with $\delta=$\,0.3--0.8 and $n_{1}=$\,0--5\,cm$^{-3}$.
Our estimate was carried out as follows. For each $\{n_{1},\delta\}$ parameter combination, we determined $K_{0}/L$ 
from fits to the existing \BC{} ratio data. Then we defined the true model using $\{n_{1}, \tau_{\rm snr}\}$ as 
source parameters and $\{\delta, K_{0}/L\}$ as transport parameters. 
From the true model, we generated the \AMS-02 mock data for a given
exposure factor, $\mathcal{F}$, and a 5\,\% systematic error. 
Thus, we re-fit the mock \BC{} ratio, leaving $K_{0}/L$ and $\delta$ as free parameters, 
within both the \textit{reference model} and the true SNR scenario (fragmentation 
specified by $n_{1}$). Finally, we estimate the \AMS-02 discrimination probability for the two models. 
The shaded areas of Fig.\,\ref{Fig::ccAMSProjection} indicate the parameter region where the \AMS-02 
discrimination succeeds at 95\%\,CL for $\mathcal{F}=$\,12, 36, and 150\,m$^{2}$\,sr\,day.
The figure shows that \AMS-02 is sensitive to a large region of the parameter space,
except for small $n_{1}$ values (small secondary SNR component) and/or small $\delta$ values 
(hard ISM component), when the intensity of the secondary source component is too weak to induce appreciable 
biases in the propagation parameters. This is also the case for Kolmogorov-like diffusion ($\delta=$\,1/3), 
which, however, is disfavored by our analysis of the real data. 
These considerations can be much strengthened if one considers the independent 
constraints that may be brought by other \AMS-02 data such as, for example,
the ratios $\overline{\textsf{p}}/\textsf{p}$, \Li/\C{}, \F/\Ne, or \Ti/\Fe{}.
In summary, our  estimates show that \AMS-02 performs wel in determining the CR transport 
parameters, providing tight constraints and considerable progress in understanding the CR acceleration 
and propagation processes. 

\section{Conclusions }                
\label{Sec::ConclusionAndDiscussion}  

We have studied the CR propagation physics under the scenarios where secondary nuclei 
can be produced or re-accelerated by Galactic sources. We have considered the processes of 
secondary productions inside SNRs and re-acceleration of background CRs in strong shocks.
The two mechanisms complement to each other and depend on the properties of the local ISM around 
the expanding remnants. The secondary production in SNRs is significant for dense background 
media, $n_{1}\gtrsim 1$\,cm$^{-3}$, while the amount of re-accelerated CRs is relevant to SNRs 
expanding into rarefied media, $n_{1}\lesssim 0.1$\,cm$^{-3}$. 
The consequence of both mechanisms is a slight flattening of the secondary-to-primary ratios 
at energies above $\sim$\,100\,GeV\,nucleon$^{-1}$. 
For the \BC{} ratio, the increase may be a factor of a few at 1 TeV \,nucleon$^{-1}$
and reach an order of magnitude at 10 TeV \,nucleon$^{-1}$.
Modeling these effects introduces an additional 
degeneracy between the source and the transport parameters. 
The diffusion coefficient index $\delta$ determined from the \BC{} 
ratio measurements above $\sim$\,10\,GeV\,nucleon$^{-1}$, 
was found to be underestimated by a factor of $\gtrsim$\,15\,\% if the
underlying model did not account 
for the hadronic production in SNRs with $n_{1}\gtrsim$\,2\,cm$^{-3}$ or for re-acceleration with 
$n_{1}\lesssim$\,0.02\,cm$^{-3}$. Nonetheless, the current uncertainty in $\delta$ is much larger
as the existing data suffer for a lack of precision at $E>10$\,GeV\,nucleon$^{-1}$.
We have shown that this degeneracy may be at least partially broken with data collected by 
high precision experiments such as \AMS-02.
Were propagation in the Galaxy to be described by a Kolmogorov spectrum ($\delta$=0.33), 
it would not be misunderstood with possible source effects described in this work, 
because these are expected to produce small distortions in the hard \BC{} ratio. 
On the other hand, we have shown that for $\delta \sim 0.4-0.8$ an \AMS-02 like experiment
will be able to discriminate pure propagation trends from a source contribution. 
Data around TeV \,nucleon$^{-1}$ energies will be clue at this aim. 
Systematic errors that can be contained to the $\sim 10 \%$ level will not prevent a clear discrimination
between the \textit{reference model} and the scenarios with SNR
components of secondary CRs. 

Data from single elements and antiprotons 
will help to identify the possible effects studied in this research. 
Closer inspections, including the revision of the role of convection and diffusive reacceleration,
will require more data at high energies, which may be released soon by a number of ongoing experiments. 
The long-duration balloon projects CREAM and TRACER, and the space missions \AMS-02 and PAMELA are 
currently operating with unprecedented sensitivities and energy ranges. 
Their data will provide valuable pieces of information about CR acceleration and propagation physics.

\begin{acknowledgements}
We warmly thank B. Bertucci and P. D. Serpico for a careful reading of the manuscript.
N.T. acknowledges the support of Italian Space Agency under contract ASI-INFN I/075/09/0.
\end{acknowledgements}


\end{document}